\documentclass[pra,onecolumn,floatfix,a4paper,superscriptaddress]{revtex4}
\usepackage{bm,color,graphicx,amsmath,txfonts}
\usepackage{bm}
\usepackage[colorlinks, citecolor=blue,linkcolor=blue]{hyperref}

\usepackage{float}

\begin{document}

\title{Entanglement of Formation, Gaussian Interferometric Power and Monogamy of Quantum Steering in Exciton--Cavity Optomechanical System}

\author{Hamza Harraf}
\affiliation{LPHE-Modeling and Simulation, Faculty of Sciences, Mohammed V University in Rabat, Rabat, Morocco}	
\author{Mohamed Amazioug} 
\thanks{m.amazioug@uiz.ac.ma}
\affiliation{LPTHE-Department of Physics, Faculty of sciences, Ibnou Zohr University, Agadir, Morocco}
\author{Rachid Ahl Laamara}
\affiliation{LPHE-Modeling and Simulation, Faculty of Sciences, Mohammed V University in Rabat, Rabat, Morocco}	
\affiliation{Centre of Physics and Mathematics, CPM, Faculty of Sciences, Mohammed V University in Rabat, Rabat, Morocco}

\begin{abstract}

Unlike classical correlations, quantum steering obeys strict monogamy relations, fundamentally limiting how strongly a single party can simultaneously steer multiple distant observers. In this work, we theoretically investigate quantum correlations in a hybrid exciton--optomechanical (EOM) system comprising excitonic, photonic, and mechanical modes. Within the Gaussian-state formalism, we employ the entanglement of formation (Eof) and Gaussian interferometric power (GIP) to quantify bipartite entanglement and quantum correlations even beyond entanglement, respectively. We also investigate Gaussian steering among the three modes in both steady-state and dynamical regimes. In particular, we establish and verify CKW-type monogamy inequalities for Gaussian steering, demonstrating that the distribution of bipartite steering among the three modes can be controlled through the system parameters. Furthermore, we examine the effects of exciton-drive detuning, input power, thermal noise, and temporal evolution on the generation and distribution of quantum correlations. Our results reveal that the Eof, GIP, and Gaussian steering exhibit distinct responses to variations in these parameters, highlighting the different sensitivities of these quantum resources to exciton-drive detuning, thermal fluctuations, power and system dynamics.
\end{abstract}
\maketitle

\section{Introduction}

An exciton is a bound electron--hole quasiparticle that emerges from the Coulomb-mediated interaction between an electron and a hole, exhibiting an effective bosonic character. Excitons constitute fundamental elementary excitations in semiconductors and other condensed-matter systems, where they can couple strongly to electromagnetic fields through their electric dipole moment~\cite{Keeling07,Deng10}. When the exciton-photon interaction becomes sufficiently strong to overcome the relevant dissipative processes, the bare excitonic and photonic modes hybridize, giving rise to new quasiparticles known as exciton polaritons~\cite{Hopfield58,Harraf2026exciton}. This strong-coupling regime was first experimentally elucidated in semiconductor quantum-well (QW) microcavities, where the confinement of the electromagnetic field substantially enhances the light-matter interaction~\cite{Weisbuch92,Houdre94}. The finding polaritons inherit properties from both their matter and photonic constituents, providing an experimentally accessible platform in which quantum-optical and many-body phenomena can be studied simultaneously. The realization and control of exciton-photon strong coupling have subsequently been extended to a wide range of photonic architectures and material platforms. More recently, we've focused on the exciton polariton in a variety of microcavities, including micropillar cavities, photonic-crystal slabs, whispering-gallery microcavities \cite{Sun08}, and others that have also led to greater localization and modification of the polariton dispersion \cite{Khitrova06, Sun08}. The strong coupling has been shown for the range of material systems, including organic semiconductors\cite{Lidzey98,Kena-Cohen08}, wide-bandgap semiconductors\cite{Sun08}, and perovskite based materials\cite{Su21}. strong coupling was demonstrated for variety kinds of material systems, such as organic semiconductors\cite{Lidzey98,Kena-Cohen08}, wide-bandgap semiconductors\cite{Sun08} and perovskite based materials\cite{Su21}. The emerging technologies give us the ability to engineer the polaritonic system's controllable coherence, spin properties, dissipation and nonlinearities. This have been giving us room to develop some new quantum and optoelectronic technologies. This hybrid light matter system therefore presents an excellent platform for exploring collective quantum effects and for the creation of a device that combines both optical coherences and material nonlinearities. The photon aspect allows them to be coupled and controlled through optical fields with great efficiency, while they borrow the inherent material nonlinearities and access strong interactions from the excitonic component \cite{Keeling07}. 
This hybrid light matter state is therefore a promising platform for investigating collective quantum phenomena and realizing a device with both optical coherence and matter nonlinearities \cite{Asjad19,Agarwal09,Rabl11}. 
Several polariton-based devices making use of both features have already been reported such as spin-memory devices \cite{Cerna13}, polariton light-emitting diodes \cite{Tsintzos08, Chakraborty19} and polariton transistors  \cite{Ballarini13, Zasedatelev19}.\\
 
Quantum steering, originating from the seminal Einstein--Podolsky--Rosen (EPR) paradox and Schrödinger's subsequent formulation of the steering phenomenon, represents a distinct form of quantum correlation that occupies an intermediate position between quantum entanglement and Bell nonlocality \cite{Reid2013,Milne2014}. EPR steering has been extensively investigated and demonstrated in a variety of physical platforms, including atom--mechanical systems \cite{intro10}, optomechanical systems \cite{He14,intro12,Amazioug2018,Amazioug2019,Mazaheri2020,Agarwal13,Wang2015,Groblasher09, Chan11, Amazioug2021, Amazouig2020,Olivares03,Amaziuog2020,Agarwa2016,Sete2011}, and waveguide-coupled architectures \cite{intro13,Zia2026}. More recently, one-way steering has been demonstrated in qubit optomagnonic cavities \cite{Qhamza2026, Chabar2026}, while cavity magnomechanical systems have emerged as promising platforms for the generation and manipulation of quantum correlations \cite{Hamza2025}. Physically, quantum steering describes a situation in which local measurements performed by one party can remotely affect the conditional state assigned to another spatially separated subsystem, in a manner that cannot be accounted for by a local hidden-state model \cite{Reid2013,Milne2014}. A defining characteristic of quantum steering is its intrinsic asymmetry: the ability of a subsystem \(A\) to steer another subsystem \(B\) does not, in general, imply that \(B\) can steer \(A\). This directional nature distinguishes quantum steering from symmetric forms of quantum correlations, such as entanglement \cite{Xiang2017,Ji2016,Hofer11,weedbrook2012,Wang13,yan2019,Teufel2011,Wang2015}. In continuous-variable systems, Gaussian quantum steering provides a particularly convenient framework for characterizing such directional correlations, since Gaussian states are completely determined by their first moments and covariance matrices \cite{Xiang2017,Yadin2018,Kim02}. For bipartite Gaussian states, the degree of steering can be quantified in terms of the symplectic eigenvalues of an appropriate conditional covariance matrix, thereby providing a quantitative measure of the steerability between the two subsystems \cite{Xiang2017}. This framework is particularly well suited to bosonic platforms, including optomechanical \cite{Abdi121,Qharraf2026,Brunelli21}, magnomechanical, and hybrid quantum systems, where quantum correlations can be generated and controlled through light--matter interactions, parametric processes, and engineered dissipation \cite{He14,Mazaheri2020}.
An important feature of multipartite quantum correlations is their **monogamous character**, which places fundamental constraints on how quantum correlations can be distributed among multiple subsystems. The concept of monogamy was initially established in the context of quantum entanglement and has subsequently played an important role in quantum information processing and quantum communication \cite{Wang2018}. In general, strong quantum correlations shared between two subsystems restrict the extent to which these subsystems can simultaneously share comparable correlations with additional parties. For multipartite systems, such restrictions are commonly expressed through monogamy inequalities, which quantify the distribution of pairwise and collective quantum correlations among the constituent subsystems \cite{Xiang2020}. However, monogamy is not a universal property of every quantum-correlation measure, and the Coffman--Kundu--Wootters (CKW) monogamy inequality is not necessarily satisfied by all measures of quantum correlations \cite{Deng2017,Huang2024}.
For continuous-variable Gaussian systems, considerable attention has therefore been devoted to identifying suitable measures that obey physically meaningful monogamy relations. In particular, it has been shown that three-mode Gaussian states satisfy a CKW-type monogamy inequality when the Gaussian contangle is employed as a measure of bipartite entanglement \cite{Xiang2017}. More generally, analogous monogamy relations can be formulated for Gaussian quantum steering, providing a useful framework for investigating how directional quantum correlations are distributed among the different modes of a multipartite system \cite{Xiang2017,Deng2017,Asjad2016,Xiang2020}. Such steering monogamy relations are especially relevant for identifying genuine multipartite quantum correlations and for understanding the constraints governing the sharing of steerability in complex continuous-variable quantum networks.

In this work, we develop a comprehensive theoretical framework to investigate Gaussian quantum steering among the exciton, optical,
and mechanical modes of a hybrid exciton--optomechanical (EOM) system~\cite{Zuo2024,Santos,EOM1,EOM2,EOM3,Fainstein20,Santos21,Santos23}.
The system forms a tripartite bosonic platform in which the exciton and cavity-photon modes are coupled through a linear
exciton-photon interaction, while the mechanical oscillator interacts dispersively with the intracavity optical field through
radiation pressure. Depending on the driving configuration and detuning conditions, the optomechanical interaction can effectively
realize either a parametric down-conversion (PDC) process or a state-swap interaction, corresponding respectively to Stokes and
anti-Stokes scattering. We consider two distinct coupling regimes for the exciton-photon
interaction. In the weak-coupling regime, the exciton and cavity-photon modes remain distinguishable, whereas in the
strong-coupling regime they hybridize to form two exciton-polariton modes. By appropriately tuning the system frequencies, we show that quantum correlations can be generated when the relevant excitonic,
photonic, or polaritonic modes are resonantly matched to the optomechanical Stokes and anti-Stokes sidebands. In particular, the
interplay between the exciton-photon coupling and the optomechanical interaction enables the generation and redistribution of quantum
correlations between the exciton and cavity-photon modes in the weak-coupling regime, and between the two polariton modes in the
strong-coupling regime. We further characterize the resulting Gaussian steering in both the
steady-state and dynamical regimes. Our analysis shows that directional steering can emerge naturally from the asymmetric
distribution of quantum correlations among the three modes, leading to one-way steering in appropriate parameter regimes. Remarkably,
Gaussian quantum steering can persist at elevated temperatures, including near room temperature, provided that the mechanical
dissipation remains sufficiently low, or equivalently that the mechanical resonator possesses an adequately high quality factor.
All in all, our work illustrates that hybrid exciton-optomechanical systems can be regarded as a rich platform for creating, manipulating and distributing directional quantum correlations in continuous-variable quantum information processing.\\

The remainder of this paper is organized as follows. In Sec.~\ref{sec2}, we introduce the exciton--optomechanical system, formulate its Hamiltonian, and derive the corresponding quantum Langevin equations. We then obtain the dynamical solutions and establish the covariance-matrix formalism required to characterize the quantum properties of the system. In Sec.~\ref{sec3}, we present the theoretical framework for quantifying quantum correlations, including the entanglement of formation, Gaussian interferometric power, and Gaussian quantum steering, together with the measures used to quantify bipartite steering and the corresponding monogamy relations. In Sec.~\ref{sec4}, we present and discuss our results for both the steady-state and dynamical regimes, highlighting the main features and physical implications of the quantum correlations and steering properties of the system. Section~\ref{sec5} is devoted to assessing the experimental feasibility of the proposed scheme by considering realistic system parameters and experimentally accessible operating conditions. Finally, in Sec.~\ref{sec6}, we summarize the main findings and present our conclusions.

\section{The exciton-optomechanics system}\label{sec2}

\begin{figure}[!h]
\centering
\begin{tabular}{ccc}
\includegraphics[scale=0.25]{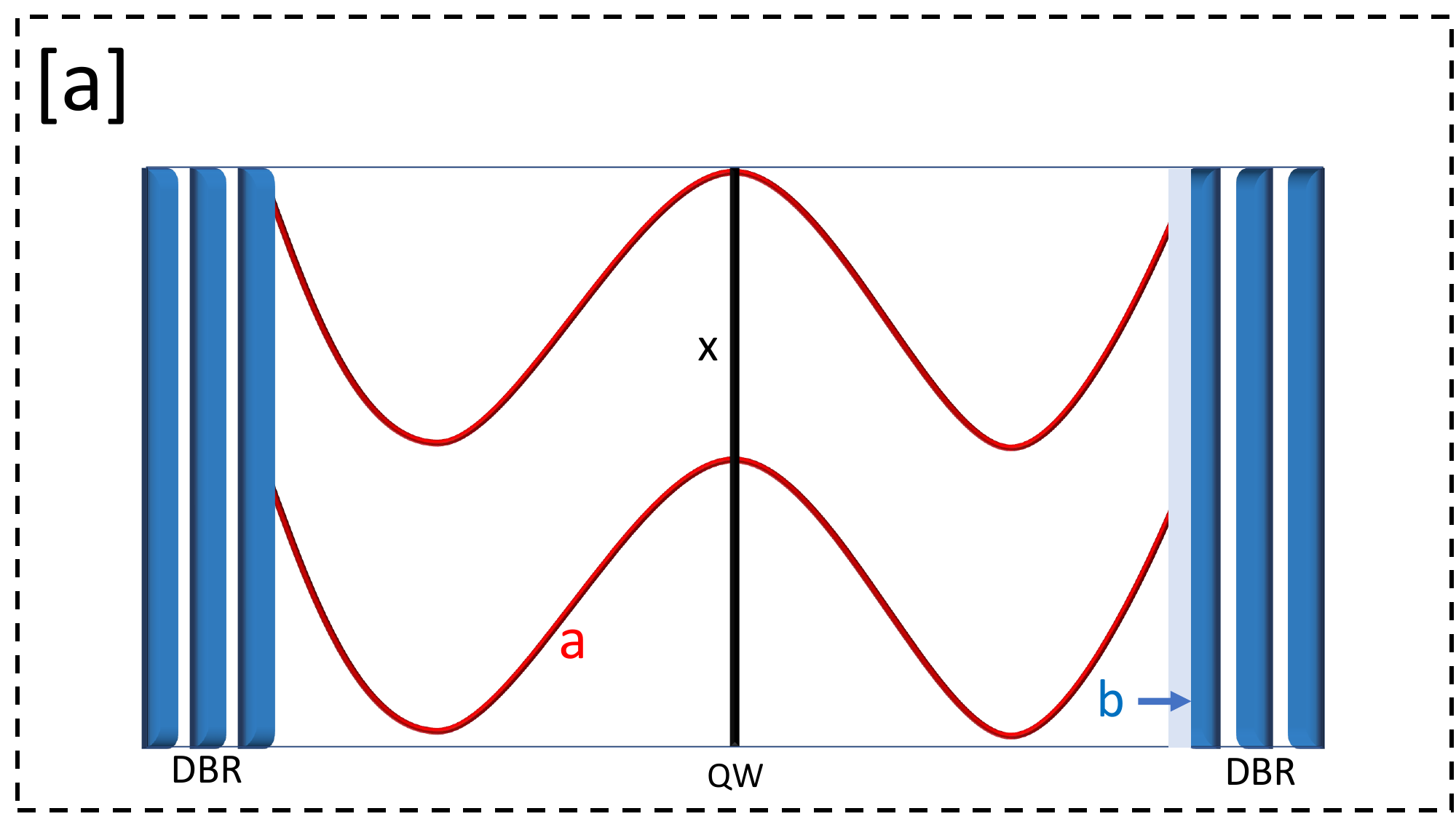} 
\hspace{0.30cm}
\includegraphics[scale=0.25]{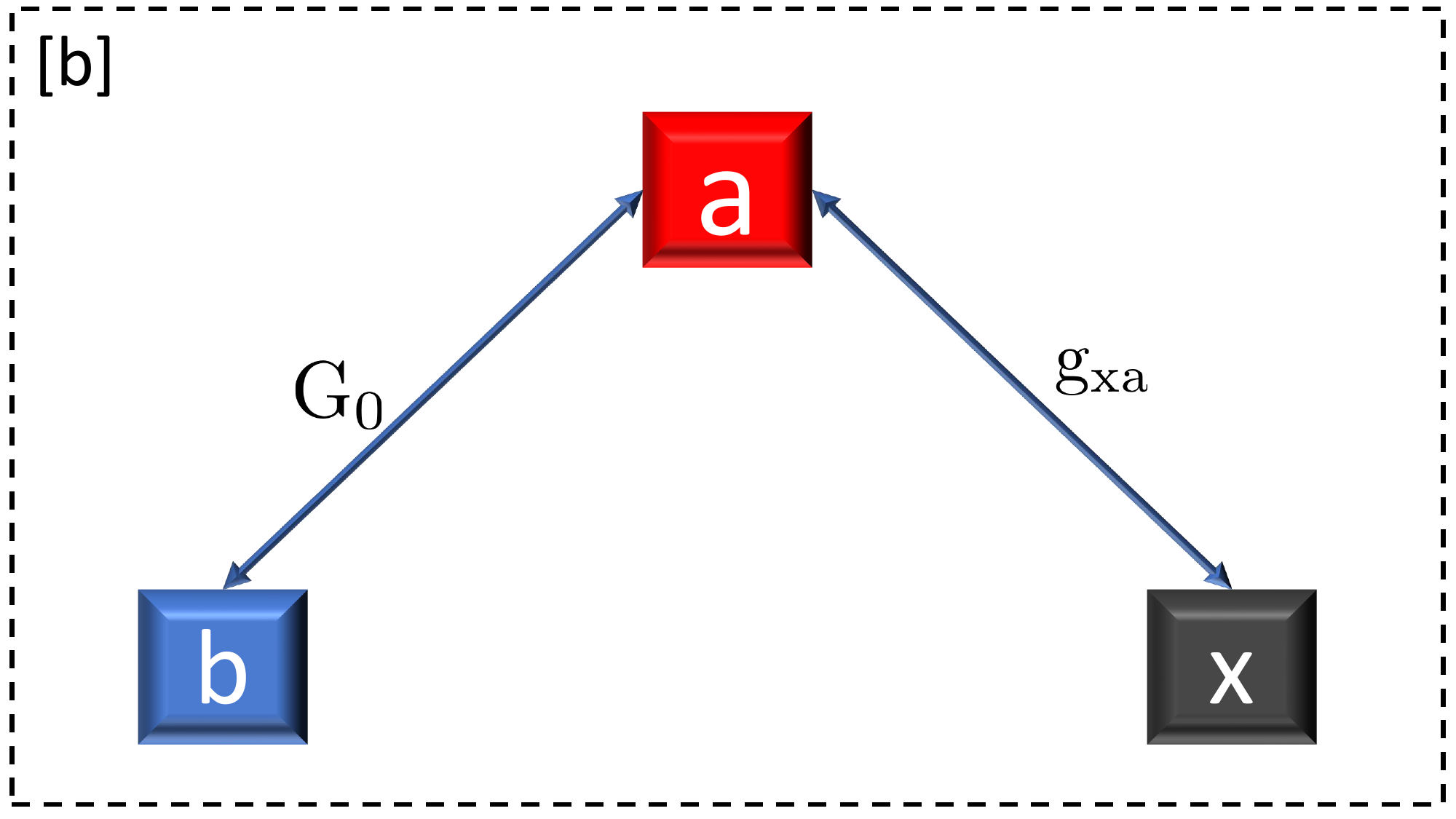}
\end{tabular}
\caption{[a] Schematic illustration of the proposed exciton--optomechanical system. A semiconductor quantum well is embedded within a microcavity bounded by two distributed Bragg reflectors (DBRs), enabling coherent coupling between the exciton ($\rm x$) and cavity mode ($\rm a$). The mechanical vibration of the DBRs introduces a mechanical degree of freedom, described by the phonon mode ($\rm b$), which couples to the intracavity optical field through radiation pressure. The sky-blue region provides a schematic illustration of the mechanical displacement for visualization purposes only and does not represent its actual amplitude or dynamical trajectory. [b] Schematic illustration of the interactions among the three subsystems. The exciton mode ($\rm x$) is coherently coupled to the cavity-photon mode ($\rm a$) with coupling strength ${\rm g}_{\rm xa}$, while the cavity-photon mode ($\rm a$) interacts with the mechanical mode ($\rm b$) through the optomechanical coupling ${\rm G_{0}}$.}
\label{fig_1}
\end{figure}

The exciton--optomechanical architecture under consideration comprises a semiconductor quantum well embedded between two mechanically compliant distributed Bragg reflectors (DBRs), forming an optomechanical microcavity, as schematically illustrated in Fig.~\ref{fig_1}[a]. Each DBR is made of alternating layers of high- and low-refractive-index material, where each of these layers is chosen such that it has an optical thickness of $\lambda/4$, where $\lambda$ is the optical wavelength. This multilayer periodic structure creates the constructive interference of the reflected optical fields and thus creates a stop band of a high optical reflection. In other words, the DBRs act as highly reflecting cavity mirrors for the optical fields of the above-mentioned frequency range, while their mechanical motion allows one to tune the cavity length.
The mechanical motion of the movable DBRs couples to the intracavity electromagnetic fields via the radiation pressure, thus creating the interface between the photonic and mechanical degrees of freedom of the EOM system~\cite{EOM1,EOM3,Perrin13}. Moreover, the quantum well is made of a thin semiconductor layer sandwiched between two thick layers of the materials that possess much wider band gap energy spectra than the semiconductor quantum well itself, thus confining the electron-hole excitations to the quantum well. The placement of the QW inside the microcavity. The Hamiltonian of the EOM model \cite{Zuo2024} is described
\begin{equation}
\label{HHH}
\mathcal{H}/\hbar
= \bm{\omega}_{\rm x} {\rm x}^\dagger {\rm x}
+ \bm{\omega}_{\rm a} {\rm a}^\dagger {\rm a}
+ \bm{\omega}_{\rm b} {\rm b}^\dagger {\rm b}
+ {\rm G}_0 {\rm a}^\dagger {\rm a} \left({\rm b}+{\rm b}^\dagger\right)
+ {\rm g}_{\rm xa}\left({\rm x}^\dagger {\rm a}+{\rm x} {\rm a}^\dagger\right)
+ i\mathcal{E}\left({\rm a}^\dagger {\rm e}^{-i\bm{\omega}_0 {\rm t}}
-{\rm a} {\rm e}^{i\bm{\omega}_0 {\rm t}}\right).
\end{equation}
Where, $\rm x$, $\rm a$, and $\rm b$ ($\rm x^\dagger$, $\rm a^\dagger$, and $\rm b^\dagger$) represent the annihilation (creation) operators associated with the exciton, cavity, and mechanical modes, respectively. They obey the standard bosonic commutation relations, $[\alpha,\alpha^\dagger]=1$, where $\alpha\in\{\rm x,a,b\}$. The corresponding resonance frequencies are denoted by $\bm{\omega}_{\rm x}$, $\bm{\omega}_{\rm a}$, and $\bm{\omega}_{\rm b}$. The coupling parameter ${\rm g}_{\rm xa}$ characterizes the exciton--photon interaction strength as shown in Fig.\ref{fig_1}[b], which depends on the position of the QW within the microcavity and can reach the strong-coupling regime. In particular, when the exciton--photon coupling exceeds both the exciton and cavity decay rates, $\gamma_{\rm x}$ and $\gamma_{\rm a}$, namely ${\rm g}_{\rm xa}>\gamma_{\rm x},\gamma_{\rm a}$, the system enters the strong-coupling regime, where exciton polaritons emerge~\cite{Weisbuch92,Houdre94}.

The single-photon optomechanical coupling strength ${\rm G}_0$ is typically weak~\cite{MA14,intro_14_5}; nevertheless, the corresponding effective optomechanical interaction can be significantly enhanced by strongly driving the microcavity with an external laser field. The last term in the Hamiltonian accounts for this coherent driving, where $\mathcal{E}=\sqrt{2{\rm P}\gamma_{\rm a}/\hbar\bm{\omega}_0}$ denotes the driving amplitude, with $\bm{\omega}_0$ and ${\rm P}$ representing the frequency and power of the incident laser field, respectively.

Taking into account the dissipation mechanisms affecting each mode, together with their associated input noise operators, the system dynamics are described by the following quantum Langevin equations (QLEs), formulated in a frame rotating at the driving frequency $\bm{\omega}_0$:
\begin{align}\label{QLExcb}
	\begin{split}
		\partial_{{\rm t}}{\rm x}=&-(i\Delta_{\rm x} + \gamma_{\rm x}) {\rm x} - i {\rm g}_{\rm xa} {\rm a} + \sqrt[]{2\gamma_{\rm x}}{\rm x}^{\rm in},  \\
		\partial_{\rm t}{\rm a}=&-(i\Delta_{\rm a} + \gamma_{\rm a}) {\rm a} - i {\rm g}_{\rm xa} {\rm x} - i {\rm G}_0 {\rm a} ( {\rm b} + {\rm b}^\dagger ) + \mathcal{E} + \sqrt[]{2\gamma_{\rm a}}{\rm a}^{\rm in}, \\
		\partial_{\rm t}{\rm b}=&-(i\bm{\omega}_{\rm b} + \gamma_{\rm b}){\rm b} - i {\rm G}_0 {\rm a}^\dagger {\rm a} + \sqrt[]{2\gamma_{\rm b}} {\rm b}^{\rm in} ,
	\end{split}
\end{align}
where $\Delta_{\rm x}=\bm{\omega}_{\rm x}-\bm{\omega}_0$ and $\Delta_{\rm a}=\bm{\omega}_{\rm a}-\bm{\omega}_0$ denote the exciton and cavity detunings, respectively, while $\gamma_{\rm b}$ represents the mechanical damping rate. The operators $\alpha^{\rm in}({\rm t})$, with $\alpha\in \{{\rm x,a,b}\}$, describe the input noise associated with the three modes. These operators have zero mean and satisfy the correlation functions~\cite{Zoller}
\begin{align}
\left\langle
\alpha^{\rm in}({\rm t})
\alpha^{{\rm in}\dagger}({\rm t}')
\right\rangle
=
\left[
{\rm N}_{\alpha}(\bm{\omega}_{\alpha})+1
\right]\delta({\rm t}-{\rm t}'),& \qquad \left\langle
\alpha^{{\rm in}\dagger}({\rm t})
\alpha^{\rm in}({\rm t}')
\right\rangle
=
{\rm N}_{\alpha}(\bm{\omega}_{\alpha})\delta({\rm t}-{\rm t}'),
\end{align}
where
$
{\rm N}_{\alpha}(\bm{\omega}_{\alpha})
=
\left[
\exp\left(
\frac{\hbar\bm{\omega}_{\alpha}}{\rm k_{\rm B}T}
\right)-1
\right]^{-1},
$
denotes the equilibrium mean thermal occupation number of mode $\alpha$, with ${\rm T}$ representing the temperature of the thermal bath and $\rm k_{\rm B}$ denotes the Boltzmann constant.\\
Under strong coherent driving, the cavity and exciton modes acquire large mean amplitudes, such that $\left|\langle {\rm a}\rangle\right|,\left|\langle {\rm x}\rangle\right|\gg1$. This permits the system dynamics to be linearized around the corresponding steady-state values by decomposing each mode operator $\alpha$ into its mean value and a quantum fluctuation, $\alpha=\langle\alpha\rangle+\delta\alpha$. By retaining only terms up to first order in the fluctuation operators and neglecting higher-order contributions, the QLEs can be separated into two independent sets: one describing the classical steady-state amplitudes and the other governing the quantum fluctuations. Accordingly, the linearized QLEs for the fluctuation operators take the form
\begin{align}
\label{QLEs}
\begin{split}
\partial_{\rm t} \delta {\rm x}
={}& -\left(i\Delta_{\rm x}+\gamma_{\rm x}\right)\delta {\rm x} -i{\rm g}_{\rm xa}\delta {\rm a} +\sqrt{2\gamma_{\rm x}}\,{\rm x}^{\rm in},\\
\partial_{\rm t}\delta {\rm a}
={}&-\left(i\tilde{\Delta}_{\rm a}+\gamma_{\rm a}\right)\delta {\rm a}-i{\rm g}_{\rm xa}\,\delta {\rm x}-{\rm G}_{\rm cb}\left(\delta {\rm b}+\delta {\rm b}^{\dagger}\right)+\sqrt{2\gamma_{\rm a}}\,{\rm a}^{\rm in},\\
\partial_{\rm t}\delta {\rm b}
={}&-\left(i\bm{\omega}_{\rm b}+\gamma_{\rm b}\right)\delta {\rm b}-\left({\rm G}_{\rm cb}\delta {\rm a}^{\dagger}-{\rm G}_{\rm cb}^{*}\delta {\rm a}\right)+\sqrt{2\gamma_{\rm b}}\,{\rm b}^{\rm in},
\end{split}
\end{align}
where $\tilde{\Delta}_{\rm a}=\Delta_{\rm a}+2{\rm G}_0\langle{\rm b}\rangle$ denotes the effective cavity drive detuning, which incorporates the frequency shift induced by the optomechanical interaction, while ${\rm G}_{\rm cb}=i{\rm G}_0\langle{\rm a}\rangle$ represents the corresponding effective optomechanical coupling strength. The steady-state classical amplitudes are given by
\begin{align}\label{stSol}
\langle{\rm x}\rangle &= \frac{-i\mathcal{E}{\rm g}_{\rm xa}}{{\rm g}_{\rm xa}^2+(i\tilde{\Delta}_{\rm a}+\gamma_{\rm a})(i\Delta_{\rm x}+\gamma_{\rm x})}, \qquad
\langle{\rm a}\rangle = \frac{\mathcal{E}(i\Delta_{\rm x}+\gamma_{\rm x})}{{\rm g}_{\rm xa}^2+(i\tilde{\Delta}_{\rm a}+\gamma_{\rm a})(i\Delta_{\rm x}+\gamma_{\rm x})}, \qquad
\langle{\rm b}\rangle = -\frac{{\rm G}_0}{{\bm\omega}_{\rm b}}|\langle{\rm a}\rangle|^2 .
\end{align}
The set of QLEs in Eq.~\eqref{QLEs} can be recast into a compact matrix representation in terms of the quadrature fluctuation operators as follows: $\delta X_{\alpha} = (\delta \alpha + \delta \alpha^\dagger)/\!\sqrt{2}$, and $\delta Y_{\alpha} = i (\delta \alpha^\dagger - \delta \alpha)/\!\sqrt{2}$, i.e.,
\begin{align}\label{uAn}
\partial_{\rm t} \psi({\rm t})= {\cal A} \, \psi({\rm t}) + \xi({\rm t}),
\end{align}
where, $\psi({\rm t})=\left[\delta\mathbf{R}_{\rm x}({\rm t}),\delta\mathbf{R}_{\rm c}({\rm t}),\delta\mathbf{R}_{\rm b}({\rm t})\right]^{\rm T}$ denotes the vector of quantum fluctuations, where $\delta\mathbf{R}_{\alpha}({\rm t})=\left[\delta X_{\alpha}({\rm t}),\delta Y_{\alpha}({\rm t})\right]^{\rm T}$ represents the quadrature fluctuations of mode $\alpha$, with $\alpha\in\{{\rm x,c,b}\}$. The corresponding input-noise vector is given by $\xi({\rm t})=\left[\sqrt{2\kappa_{\rm x}}\mathbf{R}_{\rm x}^{\rm in}({\rm t}),\sqrt{2\kappa_{\rm c}}\mathbf{R}_{\rm c}^{\rm in}({\rm t}),\sqrt{2\kappa_{\rm b}}\mathbf{R}_{\rm b}^{\rm in}({\rm t})\right]^{\rm T}$, where $\mathbf{R}_{\alpha}^{\rm in}({\rm t})=\left[X_{\alpha}^{\rm in}({\rm t}),Y_{\alpha}^{\rm in}({\rm t})\right]^{\rm T}$ denotes the vector of input-field quadratures. The operators $X_{\alpha}^{\rm in}({\rm t})$ and $Y_{\alpha}^{\rm in}({\rm t})$ are defined analogously to the fluctuation quadratures $\delta X_{\alpha}({\rm t})$ and $\delta Y_{\alpha}({\rm t})$, but in terms of the corresponding input-noise operators $\alpha^{\rm in}({\rm t})$ and $\alpha^{{\rm in}\dagger}({\rm t})$.\\
where $\mathcal{A}$ is the drift matrix governing the linearized dynamics of the system.

\begin{align}
	\cal A=\begin{pmatrix}
		-\gamma_{\rm x} & {\Delta}_{\rm x} & 0 & {\rm g_{\rm xa}} & 0 & 0 \\
		-{\Delta}_{\rm x} & -\gamma_{\rm x} & -{\rm g_{\rm xa}} & 0 & 0 & 0\\
		0 & {\rm g_{\rm xa}} & -\gamma_{\rm a} & \tilde{\Delta}_{\rm a} & - 2 {\rm Re}\left[{\rm G}_{\rm cb}\right] &0 \\
		-{\rm g_{\rm xa}} & 0 & -\tilde{\Delta}_{\rm a} & -\gamma_{\rm a} & - 2 {\rm Im}\left[{\rm G}_{\rm cb}\right] & 0\\
		0 & 0 & 0 & 0 & -\gamma_{\rm b} & {\bm\omega}_{\rm b} \\
		0 & 0 & - 2 {\rm Im}\left[{\rm G}_{\rm cb}\right] & 2 {\rm Re}\left[{\rm G}_{\rm cb}\right] & -{\bm\omega}_b & -\gamma_{\rm b}\\
	\end{pmatrix}.
\end{align}
Because of the linearity of the dynamics and the fact that the quantum noise operators are Gaussian, the steady-state distribution of the quadrature fluctuations is a Gaussian state for three modes, fully determined by the $6 \times 6$ covariance matrix (CM), ${\cal V}$, with its entries $\mathcal{V}_{ij}=\frac{1}{2} \langle\psi_i({\rm t})\psi_j({\rm t}^\prime) + \psi_j({\rm t}^\prime)\psi_i({\rm t}) \rangle$ with $(i,j\in\{1,2,\dots,6\})$. The dynamical covariance matrix can be found using the solution to the Lyapunov equation associated with the dynamical system~\cite{Parks,Vitali07,Mauro07}
\begin{align}
\label{Cov}
	\begin{split}
	\partial_{\rm t}{\cal V}={\cal A} {\cal V} + {\cal V} {\cal A}^{\rm T}+{\bm D},
	\end{split}
\end{align}
where 
\begin{equation}
{\bm D}
=
\begin{pmatrix}
\gamma_{\rm x}(2{\rm N}_{\rm x}+1)\mathbb{I}_2 & 0 & 0\\
0 & \gamma_{\rm a}(2{\rm N}_{\rm a}+1)\mathbb{I}_2 & 0\\
0 & 0 & \gamma_{\rm b}(2{\rm N}_{\rm b}+1)\mathbb{I}_{2}
\end{pmatrix},
\end{equation}
where ${\bm D}$ denotes the diffusion matrix, whose elements are defined through the symmetrized noise correlation function,
$
{\bm D}_{ij}\,
=
\frac{1}{2\delta(t-t')}
\left\langle
\xi_i(t)\xi_j(t')
+
\xi_j(t')\xi_i(t)
\right\rangle,
$
where $\xi_i(t)$ represents the $i$th component of the corresponding input-noise vector.

\section{Quantum correlation}\label{sec3}
The covariance matrix $\mathcal{V}_{\bm{BA}}$ describing the correlations between the two modes $\bm B$ and $\bm A$ can be written in block form as
\begin{equation}
\mathcal{V}_{\bm{BA}} =
\begin{pmatrix}
\mathcal{B} & \mathcal{C} \\
\mathcal{C}^{T} & \mathcal{A}
\end{pmatrix},
\label{eq:covariance_XY}
\end{equation}
where the $2\times 2$ submatrices are given by $ \mathcal{B}$, and $\mathcal{A}$ which describe the local autocorrelations of modes $\bm B$ and $\bm A$, respectively. The cross-correlation matrix is given by $\mathcal{C}$, and characterizes the intermodal correlations between the two modes.
\subsection{Entanglement of formation}
The Gaussian state of the system is completely described by its covariance matrix ${\cal V}$, defined in Eq.~(\ref{Cov}). Its essential properties can be characterized in terms of the symplectic eigenvalues $\nu_{\pm}$ \cite{Adesso2005,Olivares2012}, which are expressed as follows:
\begin{equation}
\nu_{\pm} = \sqrt{\frac{\sigma \pm \sqrt{\sigma^2 - 4\det({\cal V})}}{2}},
\end{equation}
with \(\sigma = {\cal B}^2 + {\cal A}^2 - 2{\cal C}^2\). The entanglement of formation (Eof) can subsequently be quantified in terms of the smallest symplectic eigenvalue $\nu_-$ \cite{Marian2008}, and is given by
\begin{equation}
{\rm E}^{\rm F} = \max\{0, {\rm h}(2\nu_-)\},
\end{equation}
with \({\rm h}(y) = \frac{(1+y)^2}{4y} \log_2 \left[ \frac{(1+y)^2}{4y} \right] - \frac{(1-y)^2}{4y} \log_2 \left[ \frac{(1-y)^2}{4y} \right]\).
The Eof quantifies the average number of maximally entangled pure-state pairs (EPR pairs) required to prepare a given quantum state using local operations and classical communication (LOCC).

\subsection{Gaussain Interfometrc power}
In this section, we investigate the quantum correlations within the hybrid exciton--optomechanical system using the Gaussian interferometric power (GIP) \cite{Adesso04,chabar2024}. In the preceding section, the entanglement between the excitonic and optical modes was examined through the positive partial transpose (PPT) criterion, which enabled us to identify the parameter regimes in which the corresponding bipartite Gaussian state is entangled. Nevertheless, the absence of entanglement does not necessarily indicate that the state is devoid of quantum correlations, as nonclassical correlations may persist even in the separable regime. To capture these correlations, we employ the GIP, which provides a suitable quantifier of quantum correlations beyond entanglement and is directly related to the precision of quantum parameter estimation \cite{Adesso04}. In particular, we evaluate the GIP with respect to the optical mode, considered as the probing subsystem, while the excitonic mode acts as the other party of the bipartite system. This choice allows us to characterize the quantum correlations encoded in the exciton--photon state and investigate their potential relevance to quantum metrology. Based on the covariance matrix defined in Eq.~\eqref{eq:covariance_XY}, the GIP can be expressed explicitly as \cite{Adesso04}

\begin{equation}
  \mathcal{I}=\frac{X+\sqrt{X^{2}+YZ}}{2Y},
  \label{eq:GIP_general}
\end{equation}
where
\begin{align}
  X &= (\alpha+\gamma)(1+\beta+\gamma-D)-D^{2}, \nonumber\\
  Y &= (D-1)(1+\alpha+\beta+2\gamma+D), \nonumber\\
  Z &= (\alpha+D)(\alpha\beta-D)+\gamma(2\alpha+\gamma)(1+\beta),
  \nonumber
\end{align}
with
  $\alpha=4\det[\bm{\mathcal{B}}(t)],$
  $\beta=4\det[\bm{\mathcal{A}}(t)],$
  $\gamma=4\det[\bm{\mathcal{C}}(t)],$
$D=16\det[\bm{\mathcal{V}}_{\bm BA}(t)].$

\subsection{Gaussian quantum steering}
Gaussian quantum steering constitutes a distinctive form of quantum correlations in continuous-variable systems, reflecting the ability of one subsystem to remotely influence the conditional quantum state of another through local measurements. For a bipartite Gaussian state comprising $(m_{\bm A}+m_{\bm B})$ modes and described by the covariance matrix $\mathcal{V}_{\bm{BA}}$, the degree of Gaussian steering exerted by subsystem $\bm B$ on subsystem $\bm A$ can be quantified by the Gaussian steering measure \cite{IKogias2015}, expressed as follows:
\begin{equation}
\mathcal{G}^{\bm B\rightarrow\bm A}
\left(\mathcal{V}_{\bm{BA}}\right)
=
\max\left[
0,
-\sum_{{\rm k}:\bar{\nu}_{\rm k}^{\bm{BA}/\bm B}<1/2}
\ln\left(\bar{\nu}_{\rm k}^{\bm{BA}/\bm B}\right)
\right],
\label{eq:37}
\end{equation}
where $\bar{\nu}_{\rm k}^{\bm{BA}/\bm B}$, with ${\rm k}=1,\ldots,m_{\bm A}$, are the symplectic eigenvalues of the Schur complement associated with subsystem $\bm B$,
the corresponding Schur complement is given by
$
\bar{\mathcal{V}}_{\bm{BA}/\bm B}
=
\mathcal{\bm A}
-
\mathcal{\bm C}^{\rm T}
\mathcal{\bm B}^{-1}
\mathcal{\bm C}.
$\\
The Gaussian steerability in the opposite direction, $\mathcal{G}^{\bm A\rightarrow\bm B}$, is obtained by interchanging the roles of $\bm A$ and $\bm B$. The two directional steering measures are generally different, reflecting the intrinsically asymmetric nature of quantum steering.

To quantify this directional asymmetry, we define the steering asymmetry as
\begin{equation}
\mathcal{G}(\bm A,\bm B)=\left|\mathcal{G}^{\bm A\rightarrow\bm B}-\mathcal{G}^{\bm B\rightarrow\bm A}\right|.
\label{eq:38}
\end{equation}
This quantity vanishes when the steering is symmetric and becomes nonzero whenever the two directional steering strengths differ. Accordingly, the Gaussian steering between two subsystems can be classified into three main regimes. In the absence of steering, both directional measures vanish,
$
\mathcal{G}^{\bm A\rightarrow\bm B}
=
\mathcal{G}^{\bm B\rightarrow\bm A}
=
0.
$
One-way steering occurs when steering is possible in only one direction, for example,
$
\mathcal{G}^{\bm B\rightarrow\bm A}>0,
$ $
\mathcal{G}^{\bm A\rightarrow\bm B}=0.
$
Two-way steering is present when both directional measures are nonzero,
$
\mathcal{G}^{\bm A\rightarrow\bm B}>0,
$ $
\mathcal{G}^{\bm B\rightarrow\bm A}>0.
$
In this case, the steering can be either asymmetric, when
$\mathcal{G}^{\bm A\rightarrow\bm B}\neq
\mathcal{G}^{\bm B\rightarrow\bm A}$, or symmetric, when
$
\mathcal{G}^{\bm A\rightarrow\bm B}
=
\mathcal{G}^{\bm B\rightarrow\bm A}>0,
$
for which $\mathcal{G}(\bm A,\bm B)=0$.

Beyond pairwise steering, the distribution of Gaussian steering correlations in a multipartite system is constrained by monogamy relations. Based on the Gaussian steering condition presented in Ref.\cite{IKogias2015}, we analyze the monogamy behavior of the Gaussian steering for the tripartite continuous variable Gaussian state used in this study \cite{intro12}. In order to identify the various steering situations, there are three cases to be considered, namely (i) bipartite steering, which is when one mode steers the other and can be represented by $\gamma\to\alpha$; (ii) one versus two modes steering, which occurs when one mode steers the composite mode of the other two, and can be denoted as $\gamma\to\alpha\beta$; and (iii) two versus one modes steering, where the composite mode $\alpha\beta$ steers the mode $\gamma$ and is denoted by $\alpha\beta\to\gamma$.

The monogamy of Gaussian steering can be characterized through CKW-type inequalities \cite{Coffman,Xiang2017},
\begin{equation}
\begin{aligned}
\mathcal{G}^{\gamma\rightarrow\alpha\beta}
-
\mathcal{G}^{\gamma\rightarrow\alpha}
-
\mathcal{G}^{\gamma\rightarrow\beta}
&\geq 0,\qquad
\mathcal{G}^{\alpha\beta\rightarrow\gamma}
-
\mathcal{G}^{\alpha\rightarrow\gamma}
-
\mathcal{G}^{\beta\rightarrow\gamma}
&\geq 0.
\end{aligned}
\label{eq:39}
\end{equation}
The first inequality states that the collective steering of the pair $(\alpha,\beta)$ by subsystem $\gamma$ is at least as large as the sum of the corresponding pairwise steering contributions from $\gamma$ toward $\alpha$ and $\beta$. Similarly, the second inequality expresses the analogous constraint for the joint steering of subsystem $\gamma$ by the pair $(\alpha,\beta)$. These relations characterize the distribution of Gaussian steering correlations among the different bipartitions of the tripartite state and quantify their monogamous nature.
\section{Results and discussions} \label{sec4}
\begin{table}[htbp]
\centering
\caption{Parameters utilized in the numerical simulations \cite{Zuo2024}.}
\label{table}
\begin{tabular*}{\textwidth}{@{\extracolsep{\fill}} l c c c}
\hline\hline
Parameter & Symbol & Value & Unit \\
\hline
Mechanical frequency  & ${\bm \omega}_{\rm b}/2\pi$  & $20\times10^{9}$  & Hz \\ 
Mechanical damping rate  & $\gamma_{\rm b}/2\pi$  & $1\times10^{6}$  & Hz \\ 
Cavity decay rate  & $\gamma_{\rm a}/2\pi$  & $1\times10^{9}$  & Hz \\
Exciton decay rate  & $\gamma_{\rm x}/2\pi$  & $100\times10^{6}$  & Hz \\
Single-photon optomechanical coupling  & ${\rm G}_0/2\pi$  & $10\times10^{6}$  & Hz \\
Effective driving amplitude  & $\mathcal{E}/2\pi$  & $6\times10^{12}$  & Hz \\
Drive power  & ${\rm P}$  & $26\times10^{-3}$  & W \\
Reference frequency  & ${\bm \omega}_0/2\pi$  & $345\times10^{12}$ & Hz \\
Optomechanical coupling strength  & ${\rm g}_{\rm xa}/2\pi$  & $0.9\times10^{9}$  & Hz \\
Temperature  & ${\rm T}$  & $1$  & K \\
\hline\hline
\end{tabular*}
\end{table}
\subsection{Steady state}
\begin{figure}[ht!]
\centering
\begin{tabular}{ccc}
\includegraphics[scale=0.4]{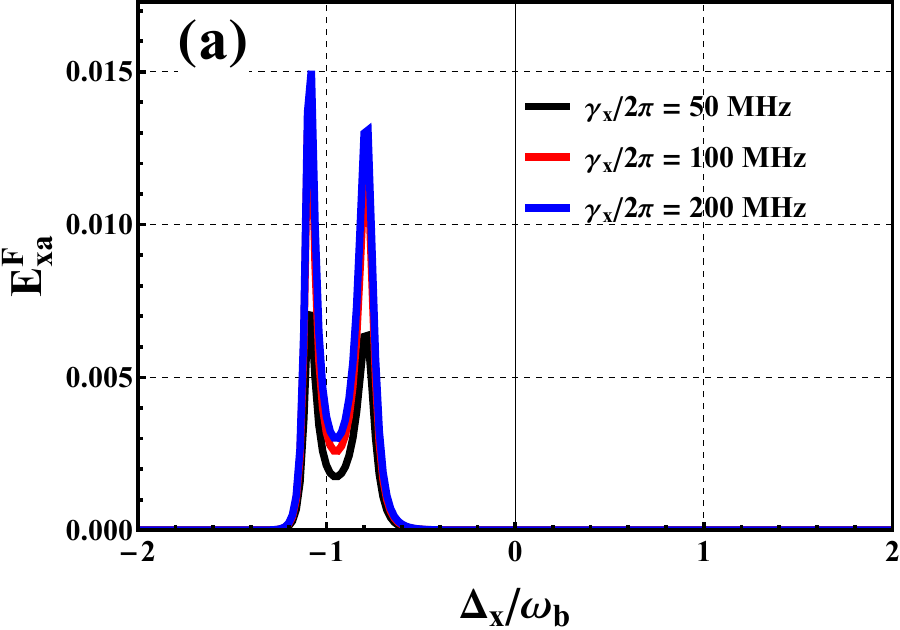} 
\includegraphics[scale=0.4]{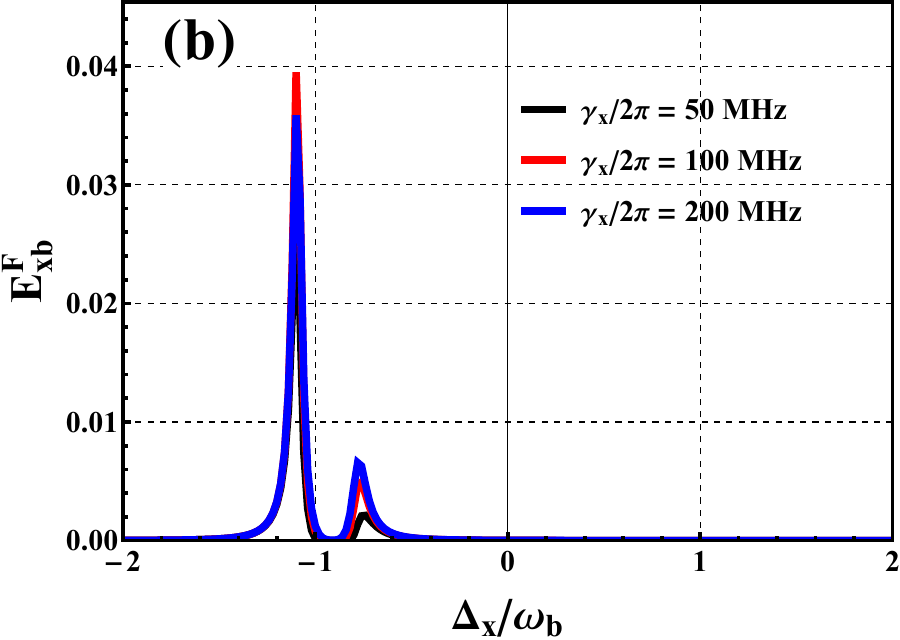}
\includegraphics[scale=0.4]{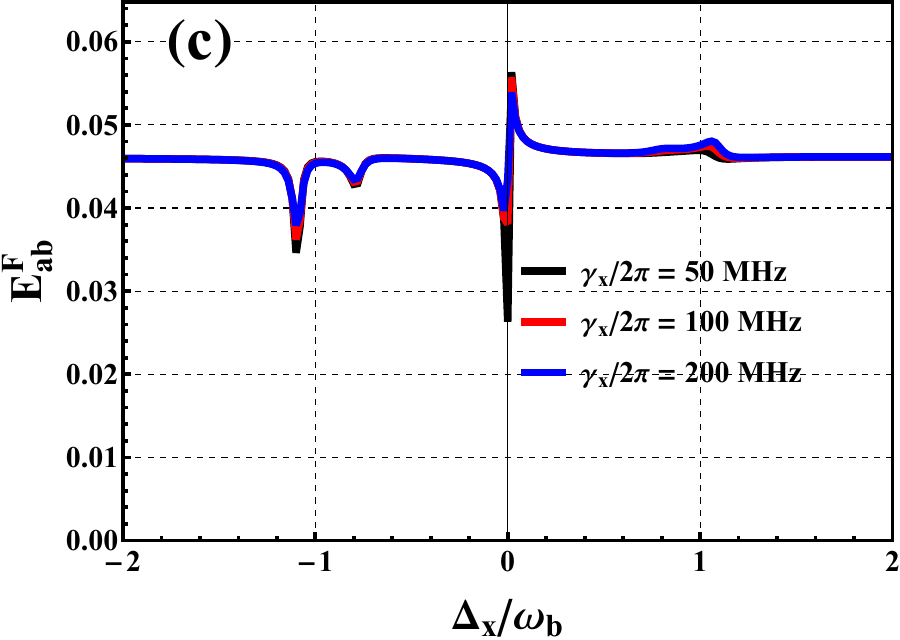} \\ 
\includegraphics[scale=0.4]{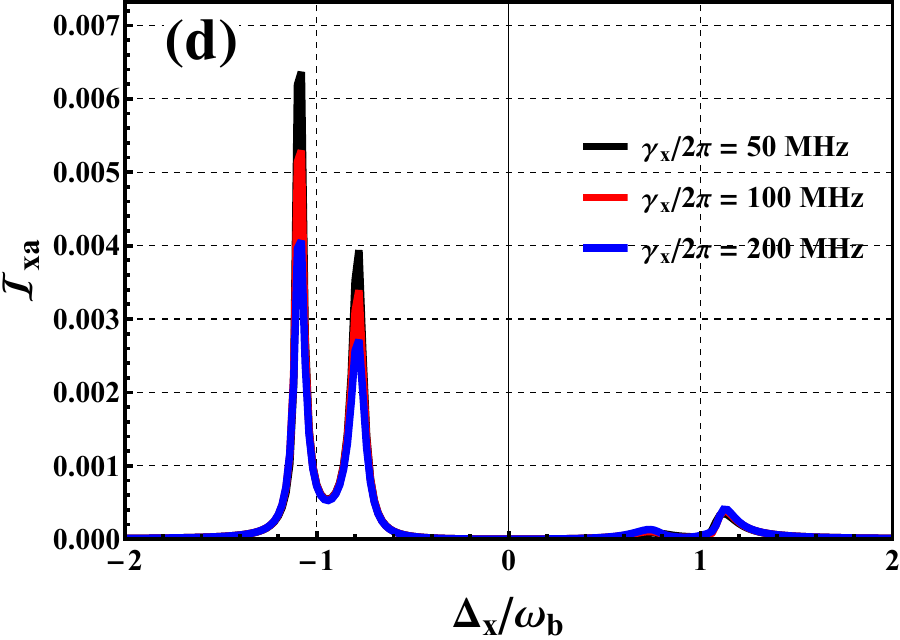} 
\includegraphics[scale=0.4]{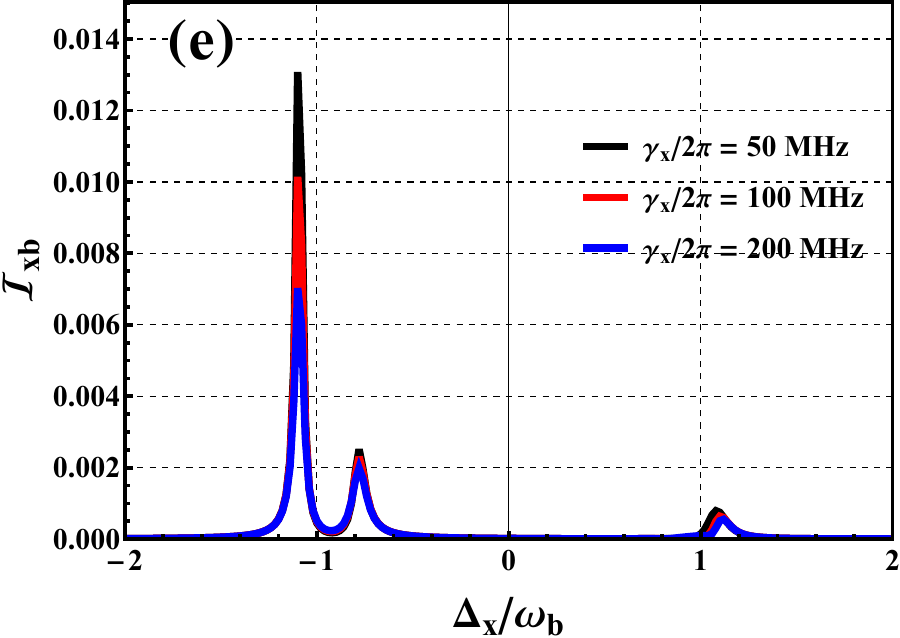} 
\includegraphics[scale=0.4]{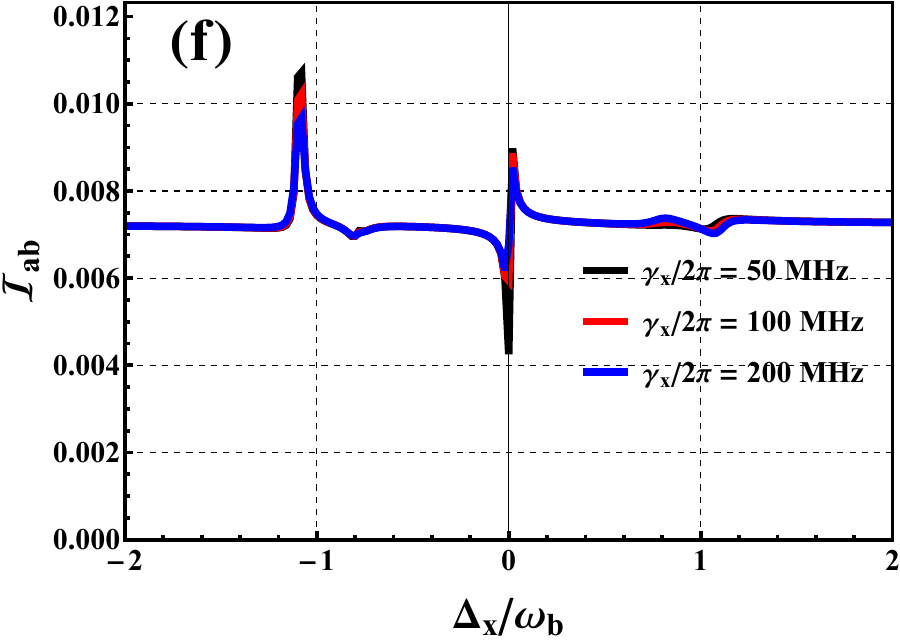} 
\end{tabular}
\caption{Plot the variation of the steady state of ((a)--(c)) the entanglement of formation  and ((d)--(f)) Gaussian interferometric power, for different bipartite subsystems: (a,d) exciton–photon (${\rm E^{F}_{xa}},{\cal{I}_{\rm xa}}$), (b,e) exciton–phonon (${\rm E^{F}_{xb}},{\cal{I}_{\rm xb}}$), and (c,f) photon–phonon modes (${\rm E^{F}_{ab}},{\cal{I}_{\rm ab}}$), as function of the exciton-drive detuning $\Delta_{\rm x}/{\bm \omega}_{\rm b}$ for various values of exiton decay rate $\gamma_{\rm x}$ with $\Delta_{\rm a}=0.9{\bm \omega}_{\rm b}$.}
\label{fig5}
\end{figure} 
In Fig.~\ref{fig5}, we investigate the steady-state behavior of the entanglement of formation and Gaussian interferometric power for three bipartite subsystems: (a,d) exciton--photon (${\rm E^{F}_{xa}},{\cal I}_{\rm xa}$), (b,e) exciton--phonon (${\rm E^{F}_{xb}},{\cal I}_{\rm xb}$), and (c,f) photon--phonon (${\rm E^{F}_{ab}},{\cal I}_{\rm ab}$), as functions of the normalized exciton-drive detuning $\Delta_{\rm x}/{\bm \omega}_{\rm b}$ for different values of the exciton decay rate $\gamma_{\rm x}$. This analysis provides insight into the role of exciton dissipation in controlling the formation of bipartite entanglement and the preservation of interferometric quantum correlations. In particular, it allows us to identify the detuning regions in which the two quantifiers exhibit similar or contrasting behaviors.

As illustrated in Fig.~\ref{fig5}(a), the Eof of the exciton--photon subsystem remains zero over an initial range of detuning values, indicating that the corresponding steady states are separable in this region. Bipartite entanglement emerges only within a relatively narrow detuning window, $\Delta_{\rm x}/{\bm \omega}_{\rm b}\in[-2.1,-1.9]$, before disappearing again as the detuning moves outside this interval. This behavior reveals the existence of a restricted parameter regime in which the effective exciton--photon interaction is sufficiently favorable to establish a nonseparable Gaussian state. Furthermore, increasing the exciton decay rate $\gamma_{\rm x}$ enhances the Eof within the entangled region. This seemingly counter intuitive behavior indicates that, for the considered parameter regime, exciton dissipation can modify the steady-state balance between coherent coupling and dissipation in a manner that favors exciton--photon entanglement. However, this enhancement should be understood as a parameter-dependent effect rather than a general consequence of dissipation. Interestingly, the corresponding GIP, shown in Fig.~\ref{fig5}(d), exhibits a distinct detuning dependence: it becomes appreciable around $\Delta_{\rm x}/{\bm \omega}_{\rm b}\approx 1.1$, where the Eof is negligible. Moreover, increasing $\gamma_{\rm x}$ suppresses the GIP, indicating that stronger exciton dissipation reduces the interferometric sensitivity of this subsystem to phase encoding. The different detuning regions in which Eof and GIP become significant demonstrate that entanglement and interferometric power probe different properties of the exciton--photon state.

A similar behavior is observed for the exciton--phonon subsystem, as shown in Fig.~\ref{fig5}(b). The Eof vanishes over an initial detuning range, becomes nonzero within $\Delta_{\rm x}/{\bm \omega}_{\rm b}\in[-2.2,-1]$, and subsequently disappears outside this interval. This finite entanglement window indicates that the generation of exciton--phonon entanglement is strongly controlled by the exciton-drive detuning, which determines the effective dynamical conditions under which quantum correlations are established between the excitonic and mechanical degrees of freedom. As in the exciton--photon case, increasing $\gamma_{\rm x}$ enhances the Eof within the entangled region, suggesting that an appropriate modification of the exciton dissipation can favor the steady-state entanglement in the considered regime. In contrast, the GIP displayed in Fig.~\ref{fig5}(e) becomes appreciable around $\Delta_{\rm x}/{\bm \omega}_{\rm b}\approx 1$, and decreases as $\gamma_{\rm x}$ increases. Thus, the exciton--phonon subsystem also exhibits a separation between the detuning regions supporting entanglement and those associated with enhanced interferometric quantum correlations. This distinction suggests that the presence of entanglement is not a necessary condition for obtaining a nonzero GIP, and that the latter can characterize interferometric usefulness even in parameter regions where the Eof vanishes.

A different and particularly interesting behavior emerges for the photon--phonon subsystem, as illustrated in Fig.~\ref{fig5}(c) and (f). Unlike the exciton-containing pairs, both the Eof and GIP remain finite throughout the considered detuning range. This persistence indicates that the photon--phonon subsystem maintains stronger and more robust quantum correlations against variations in the exciton-drive detuning. Such behavior can be attributed to the direct optomechanical interaction between the cavity field and the mechanical mode, which provides an effective pathway for establishing and maintaining correlations between these two degrees of freedom. Nevertheless, the two quantifiers exhibit opposite trends as the detuning is varied. In particular, around $\Delta_{\rm x}=-{\bm \omega}_{\rm b}$, the Eof decreases whereas the GIP increases. A similar competition is observed over the other detuning regions, where variations favoring one quantifier do not necessarily enhance the other. This opposite response highlights the fact that stronger entanglement does not automatically imply greater interferometric power. Indeed, the GIP quantifies the sensitivity of a quantum state to phase encoding and therefore depends on the structure of its covariance matrix and its response to local phase transformations, rather than solely on its degree of entanglement. Consequently, the photon--phonon subsystem can retain substantial interferometric usefulness even when its entanglement is reduced.

Overall, these results demonstrate that the exciton-drive detuning provides an effective control parameter for tailoring the steady-state quantum properties of the hybrid system. The exciton--photon and exciton--phonon subsystems exhibit restricted detuning windows for entanglement generation, whereas their GIP becomes appreciable in different detuning regions. By contrast, the photon--phonon subsystem maintains both entanglement and interferometric quantum correlations over the entire investigated detuning range, although the two measures respond differently to changes in detuning. Furthermore, the opposite dependence of Eof and GIP on the exciton decay rate in the exciton-containing subsystems highlights the nontrivial role of dissipation in determining the nature of steady-state quantum correlations. These findings demonstrate that optimizing entanglement and interferometric usefulness generally requires different operating conditions, emphasizing the importance of considering complementary quantum correlation measures when designing hybrid exciton--optomechanical systems for quantum information processing and precision sensing.

\begin{figure}[ht!]
\centering
\begin{tabular}{ccc}
\includegraphics[scale=0.39]{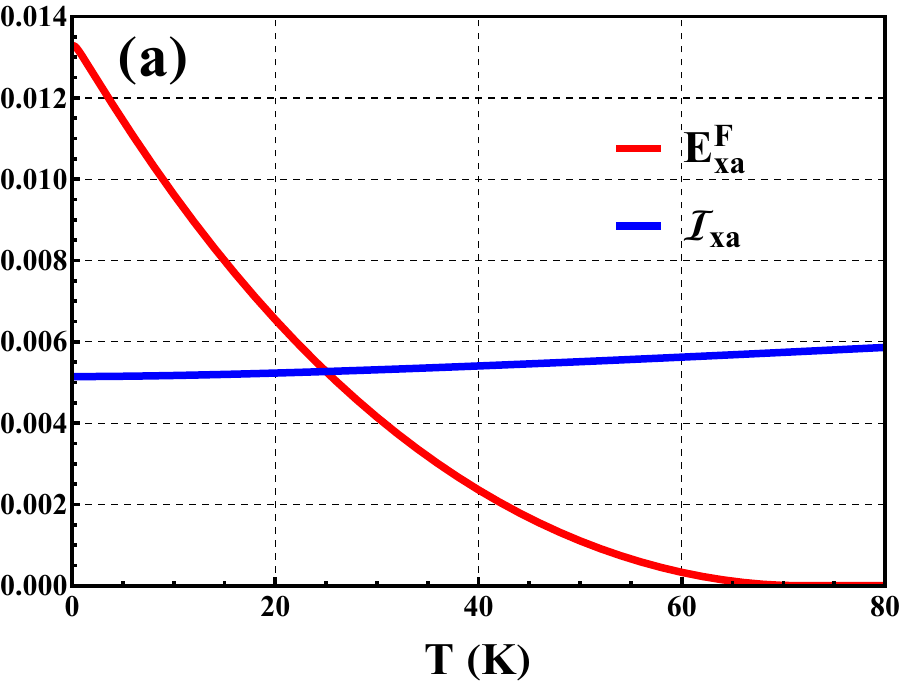} 
\hspace{0.2cm}
\includegraphics[scale=0.39]{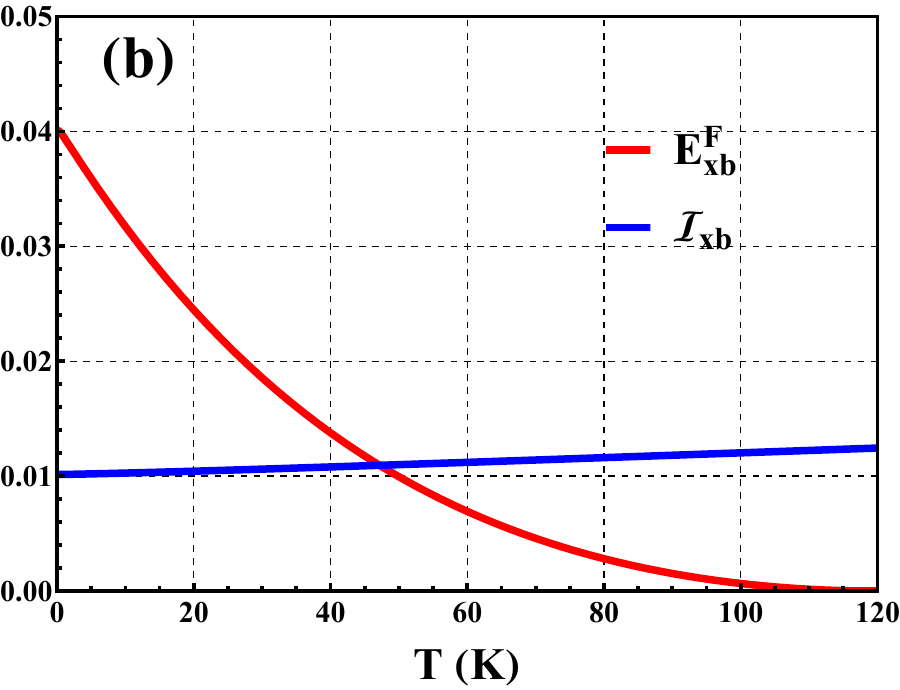}
\hspace{0.2cm}
\includegraphics[scale=0.39]{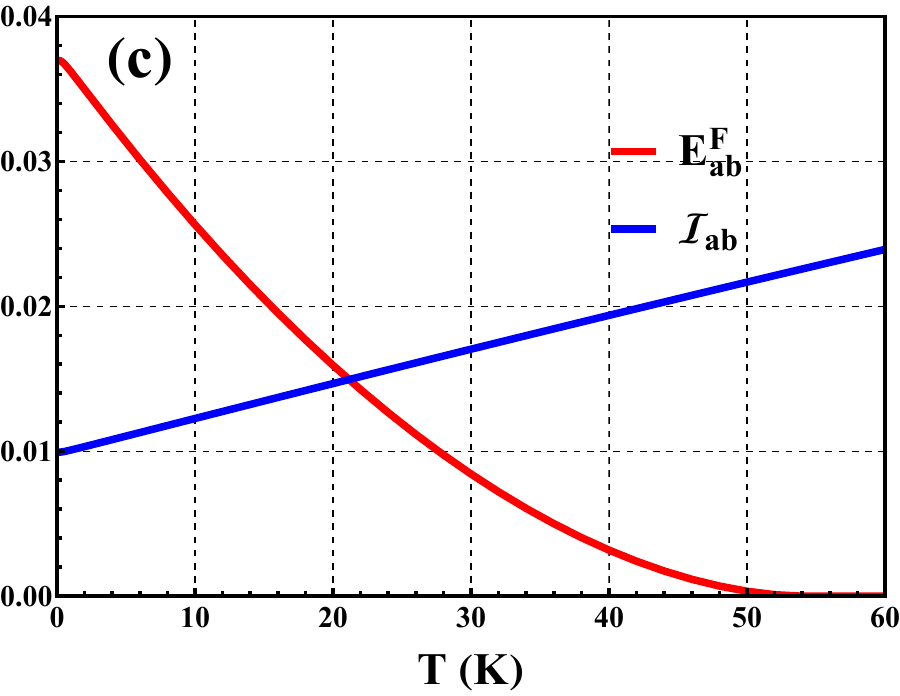} 
\end{tabular}
\caption{Plot of the steady-state of the entanglement of formation and Gaussian interferometric power as function of the temperature $\rm T$ for (a) exciton-photon mode (${\rm E^{F}_{xa}},{\cal{I}_{\rm xa}}$), (b) exciton-phonon mode (${\rm E^{F}_{xb}},{\cal{I}_{\rm xb}}$) and (c) photon-phonon mode (${\rm E^{F}_{ab}},{\cal{I}_{\rm ab}}$) with $\Delta_{\rm a}=0.9{\bm \omega}_{\rm b}$and $\Delta_{\rm x}=-1.1{\bm \omega}_{\rm b}$.}
\label{fig3}
\end{figure}

In Fig.~\ref{fig3}, we investigate the temperature dependence of the entanglement of formation  and Gaussian interferometric power for three bipartite subsystems: (a) the exciton--photon pair (${\rm E}^{\rm F}_{\rm xa},{\cal I}_{\rm xa}$), (b) the exciton--phonon pair (${\rm E}^{\rm F}_{\rm xb},{\cal I}_{\rm xb}$), and (c) the photon--phonon pair (${\rm E}^{\rm F}_{\rm ab},{\cal I}_{\rm ab}$). This analysis allows us to compare the thermal robustness of entanglement and interferometric quantum correlations across the different pairs of modes.

As shown in Fig.~\ref{fig3}(a), the Eof between the exciton and photon modes decreases with increasing temperature, indicating that thermal fluctuations progressively degrade their quantum entanglement. This behavior can be attributed to the increased thermal occupation of the mechanical environment, which introduces additional noise into the system and weakens the quantum correlations responsible for entanglement. In contrast, the GIP exhibits a slight increase with temperature and remains finite even after the Eof vanishes. This behavior highlights an important distinction between the two quantum measures: The loss of bipartite entanglement does not necessarily entail the complete disappearance of quantum correlations, as nonclassical correlations may still persist and be captured by the Gaussian interferometric power.

A qualitatively similar thermal behavior is observed for the exciton--phonon and photon--phonon subsystems, as illustrated in Figs.~\ref{fig3}(b) and \ref{fig3}(c), respectively. Nevertheless, the magnitude of the Eof differs significantly among the three pairs of modes. In particular, the exciton--phonon subsystem exhibits a larger Eof than the other two subsystems, indicating that the exciton--phonon interaction plays an important role in establishing stronger bipartite entanglement under the considered operating conditions. Furthermore, for the photon--phonon subsystem, the GIP displays an approximately linear increase with temperature. This behavior suggests that, within the investigated temperature range, thermal fluctuations can enhance the sensitivity of this subsystem to local phase encoding, as quantified by the GIP, even as its entanglement is progressively suppressed.

Overall, these results reveal distinct thermal responses of entanglement and Gaussian interferometric correlations across the three bipartite subsystems. While increasing temperature generally weakens entanglement, the GIP exhibits a slight enhancement, indicating a different sensitivity of interferometric quantum correlations to thermal fluctuations. Moreover, the comparatively larger Eof of the exciton--phonon pair identifies this subsystem as the most favorable for generating and maintaining bipartite entanglement under the considered conditions. These findings demonstrate that the thermal robustness of quantum resources depends not only on the type of quantum measure considered but also on the choice of the interacting modes. Consequently, employing both Eof and GIP provides a more comprehensive understanding of how thermal fluctuations influence the distribution and persistence of quantum correlations in hybrid exciton--optomechanical systems.

\begin{figure}[ht!]
\centering
\begin{tabular}{ccc}
\includegraphics[scale=0.4]{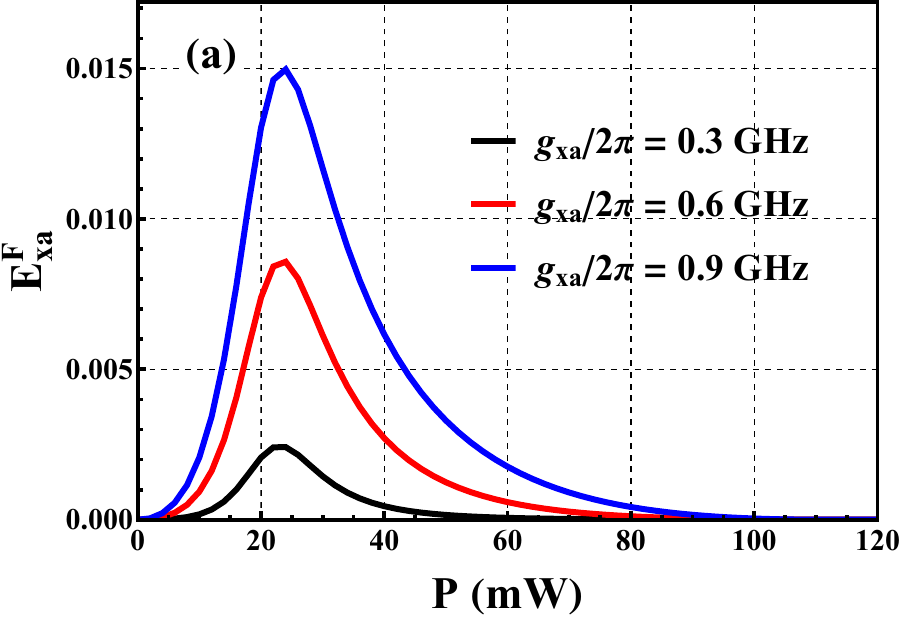}
\includegraphics[scale=0.4]{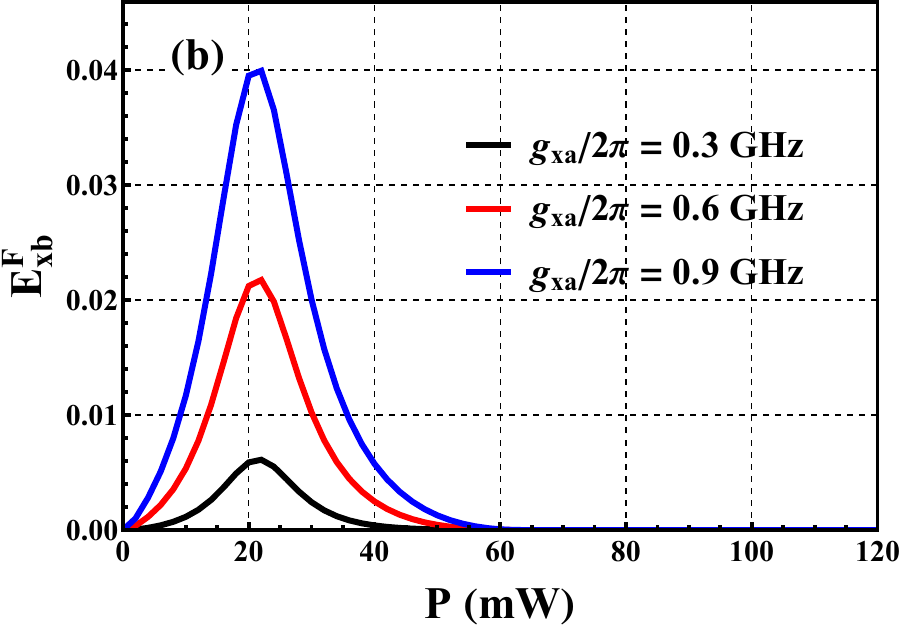}
\includegraphics[scale=0.4]{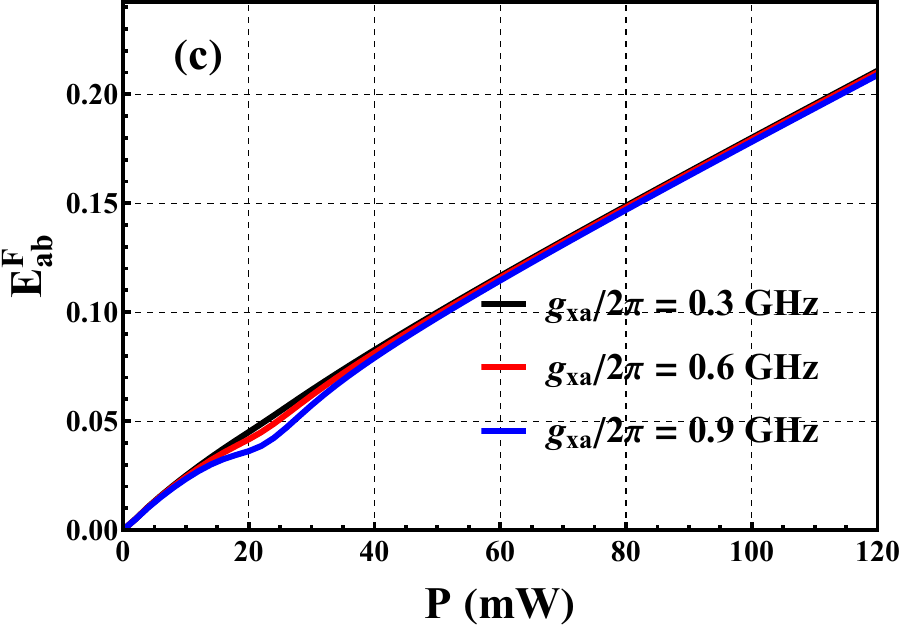} \\
\includegraphics[scale=0.4]{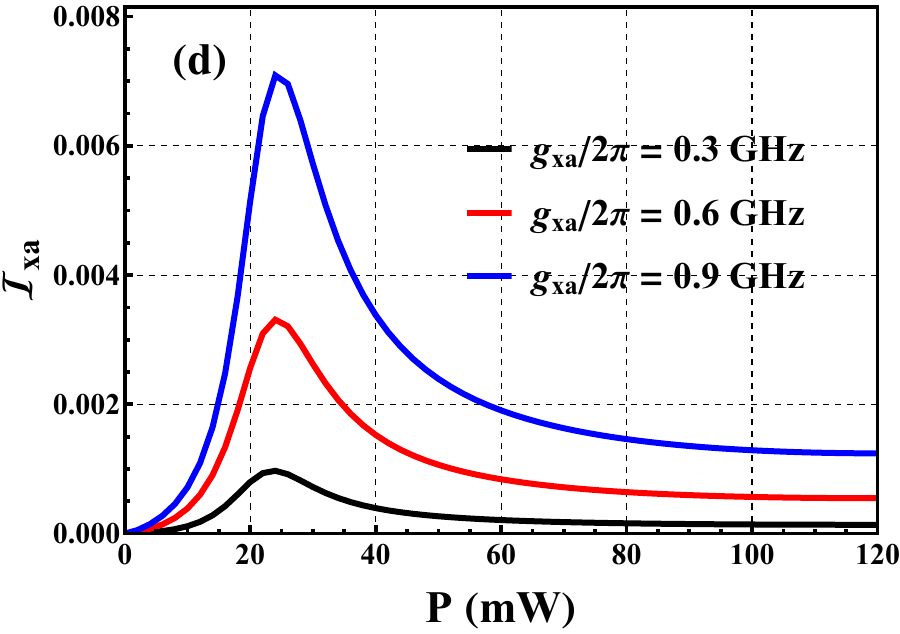} 
\includegraphics[scale=0.4]{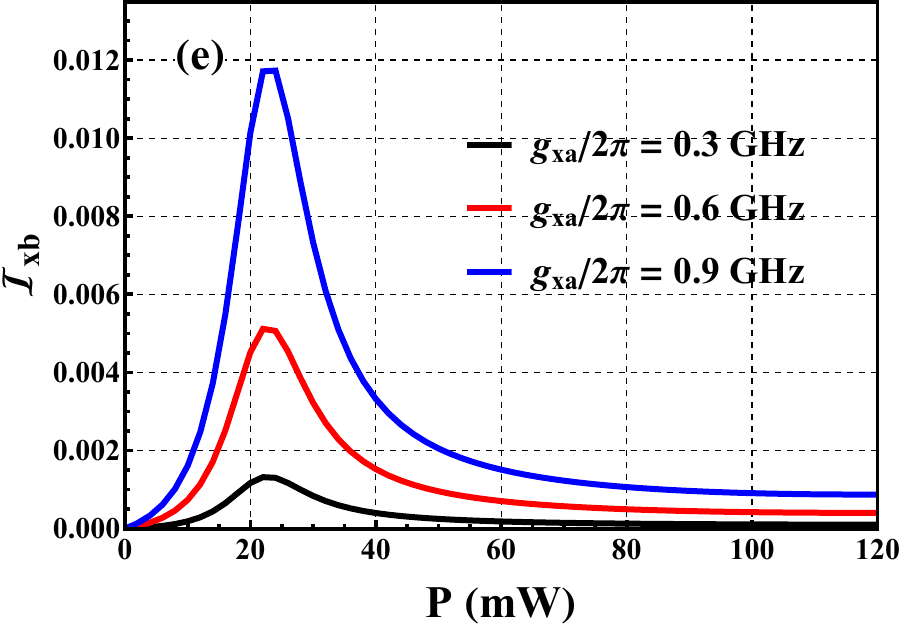} 
\includegraphics[scale=0.4]{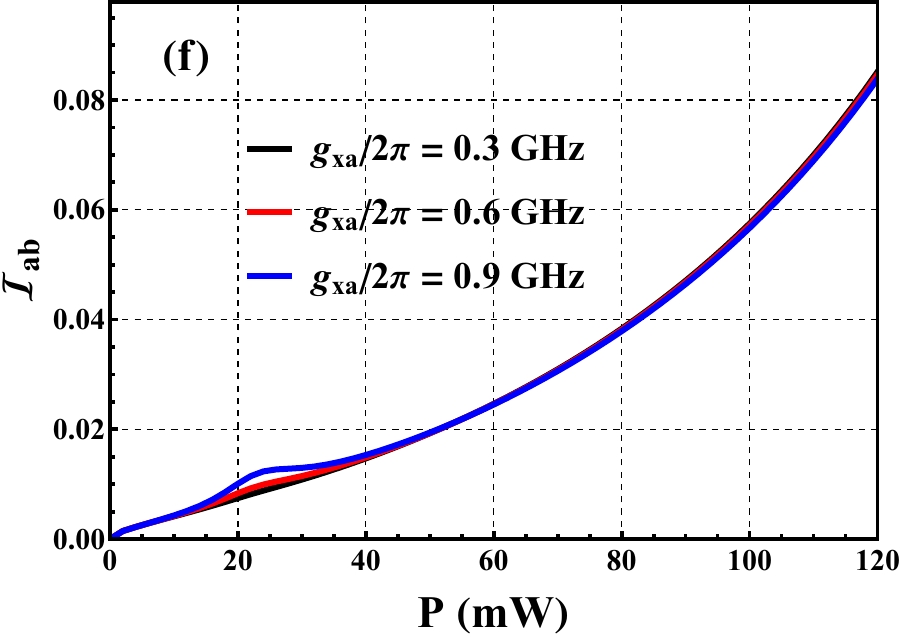} 
\end{tabular}
\caption{Plot the variation of the steady state of ((a)--(c)) the entanglement of formation  and ((d)--(f)) Gaussian interferometric power, for different bipartite subsystems: (a,d) exciton–photon (${\rm E^{F}_{xa}},{\cal{I}_{\rm xa}}$), (b,e) exciton–phonon (${\rm E^{F}_{xb}},{\cal{I}_{\rm xb}}$), and (c,f) photon–phonon modes (${\rm E^{F}_{ab}},{\cal{I}_{\rm ab}}$), as function of power $\rm P$ for various values of exiton-photon copling ${\rm g}_{\rm xa}$, with $\Delta_{\rm a}=0.9{\bm \omega}_{\rm b}$and $\Delta_{\rm x}=-1.1{\bm \omega}_{\rm b}$.}
\label{fig4}
\end{figure}

In Fig.~\ref{fig4}, we investigate the dependence of the steady-state entanglement of formation and Gaussian interferometric power  on the input power $\rm P$ for three bipartite subsystems: (a,d) exciton--photon (${\rm E}^{F}_{xa},{\cal I}_{\rm xa}$), (b,e) exciton--phonon (${\rm E}^{F}_{xb},{\cal I}_{\rm xb}$), and (c,f) photon--phonon (${\rm E}^{F}_{ab},{\cal I}_{\rm ab}$). The results are presented for different values of the exciton--photon coupling strength $g_{\rm xa}$. This analysis allows us to explore how the external driving power and the coupling between the excitonic and photonic degrees of freedom jointly influence the generation and distribution of quantum correlations in the hybrid system.

As illustrated in Fig.~\ref{fig4}(a), the Eof of the exciton--photon subsystem exhibits a nonmonotonic dependence on the input power. Initially, the entanglement increases with increasing $\rm P$, reaching a maximum around ${\rm P}=20~\rm mW$. This enhancement indicates that increasing the driving power initially strengthens the effective interaction between the exciton and photon modes, thereby promoting the generation of bipartite entanglement. Beyond this optimal driving regime, however, the Eof progressively decreases and eventually vanishes for ${\rm P}>100~\rm mW$. This behavior reveals a competition between the beneficial effect of coherent driving and the detrimental influence of dissipation and fluctuations, which become increasingly important at higher driving powers. Furthermore, increasing the exciton--photon coupling strength ${\rm g}_{\rm xa}$ enhances the Eof, demonstrating that stronger coherent coupling facilitates the establishment of nonseparable correlations between the excitonic and photonic modes. A similar nonmonotonic behavior is observed for the corresponding GIP in Fig.~\ref{fig4}(d), which also increases initially and subsequently decreases as the input power grows. However, unlike the Eof, the GIP remains finite even in the high-power regime where the entanglement vanishes. This persistence indicates that the exciton--photon subsystem can retain interferometric quantum correlations even when its bipartite entanglement is completely suppressed. Moreover, increasing ${\rm g}_{\rm xa}$ enhances the GIP, suggesting that stronger exciton--photon coupling also improves the sensitivity of this subsystem to phase encoding.

A comparable behavior is observed for the exciton--phonon subsystem, as shown in Fig.~\ref{fig4}(b). The Eof initially increases with the input power and reaches its maximum around ${\rm P}=20~\rm mW$. For larger driving powers, the entanglement decreases and eventually disappears for ${\rm P}>60~\rm mW$. The relatively lower power threshold for the disappearance of exciton--phonon entanglement, compared with the exciton--photon pair, indicates that the two subsystems exhibit different sensitivities to strong driving conditions. This difference can be associated with their distinct effective interaction pathways and their respective responses to dissipation and fluctuations. Furthermore, increasing the exciton--photon coupling strength ${\rm g}_{\rm xa}$ enhances the Eof, highlighting the role of exciton--photon hybridization in mediating quantum correlations between the excitonic and mechanical degrees of freedom. The corresponding GIP, presented in Fig.~\ref{fig4}(e), follows a similar overall power dependence, exhibiting an initial enhancement followed by a reduction at higher powers. Nevertheless, it remains nonzero beyond the power threshold at which the Eof vanishes. This result further confirms that the loss of bipartite entanglement does not necessarily imply the complete disappearance of interferometric quantum correlations. In addition, the GIP increases with ${\rm g}_{\rm xa}$, indicating that stronger exciton--photon coupling enhances the interferometric usefulness of the exciton--phonon subsystem.

In contrast, the photon--phonon subsystem exhibits a markedly different response to the input power, as illustrated in Fig.~\ref{fig4}(c). In this case, the Eof increases with increasing driving power throughout the considered parameter range, suggesting that stronger external driving progressively promotes the generation of entanglement between the optical and mechanical modes. Interestingly, around ${\rm P}\approx20~\rm mW$, the Eof decreases with increasing exciton--photon coupling strength ${\rm g}_{\rm xa}$. This opposite dependence, compared with the exciton-containing subsystems, suggests that increasing the exciton--photon coupling does not necessarily enhance entanglement in every pair of modes. Instead, it can modify the distribution of quantum correlations throughout the hybrid system, favoring some bipartite subsystems at the expense of others. Such behavior reflects the competing nature of the different interaction pathways in a multipartite system. The corresponding GIP, shown in Fig.~\ref{fig4}(f), increases with the input power, indicating that stronger driving enhances the interferometric quantum correlations of the photon--phonon pair. Moreover, around ${\rm P}\approx22~\rm mW$, the GIP also increases with increasing $g_{\rm xa}$, in contrast to the behavior of the Eof observed around ${\rm P}\approx20~\rm mW$. This difference is particularly interesting because it demonstrates that the coupling-induced redistribution of quantum correlations can affect entanglement and interferometric power in different ways. In particular, a reduction in entanglement does not necessarily correspond to a reduction in the sensitivity of the subsystem to phase encoding.

Overall, these results demonstrate that the input power and exciton--photon coupling strength provide two complementary control parameters for manipulating quantum correlations in the hybrid exciton--optomechanical system. For the exciton--photon and exciton--phonon subsystems, the Eof exhibits an optimal driving regime around $20~\rm mW$, beyond which stronger driving progressively suppresses entanglement. By contrast, the photon--phonon entanglement increases with the input power, revealing a distinct response associated with the direct optomechanical interaction. Furthermore, the persistence of GIP in the high-power regimes where Eof vanishes highlights the ability of interferometric quantum correlations to survive beyond the entanglement threshold. The different responses of the three subsystems to variations in ${\rm g}_{\rm xa}$ also emphasize the importance of considering the redistribution of quantum correlations in hybrid multipartite systems. These findings demonstrate that the optimal operating conditions for generating bipartite entanglement are not necessarily those that maximize interferometric usefulness, suggesting that the choice of driving power and coupling strength should be tailored to the specific quantum information or sensing task.

\begin{figure}[ht!]
\centering
\begin{tabular}{ccc}
\includegraphics[scale=0.45]{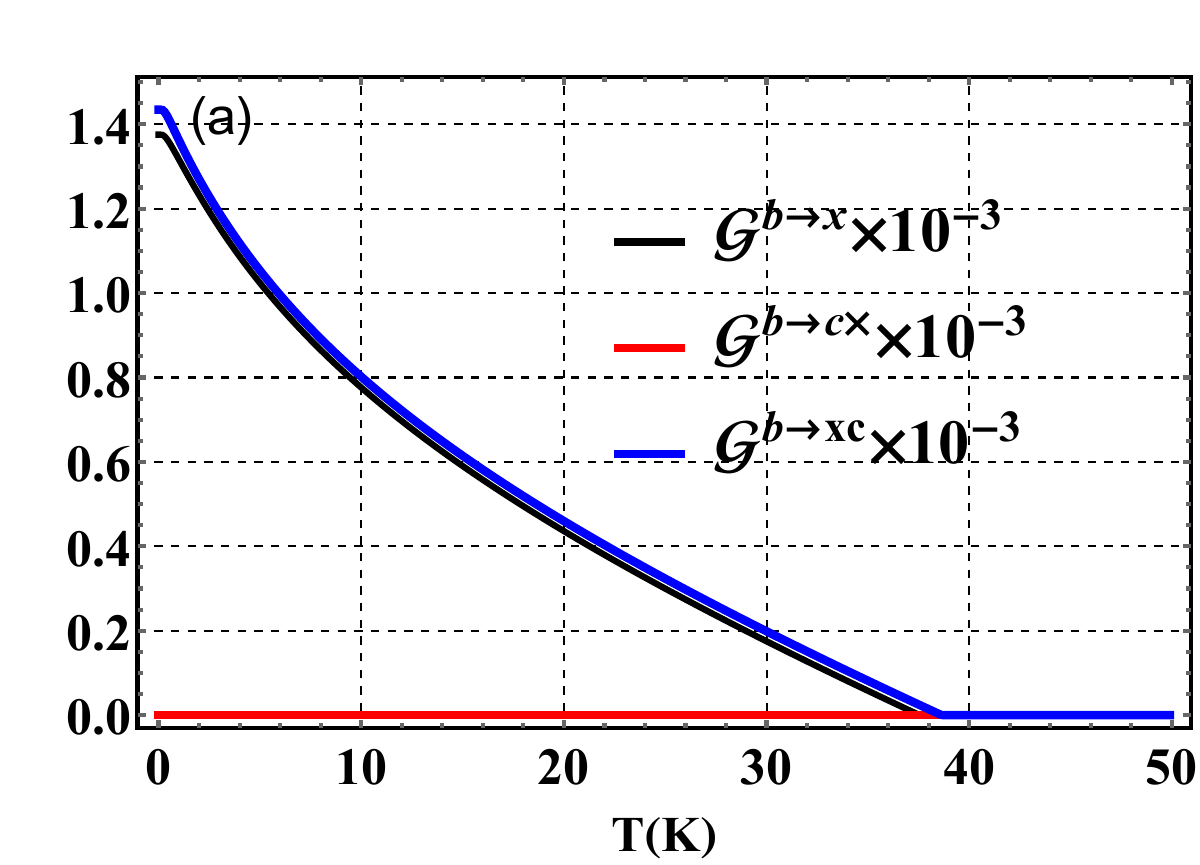} 
\includegraphics[scale=0.45]{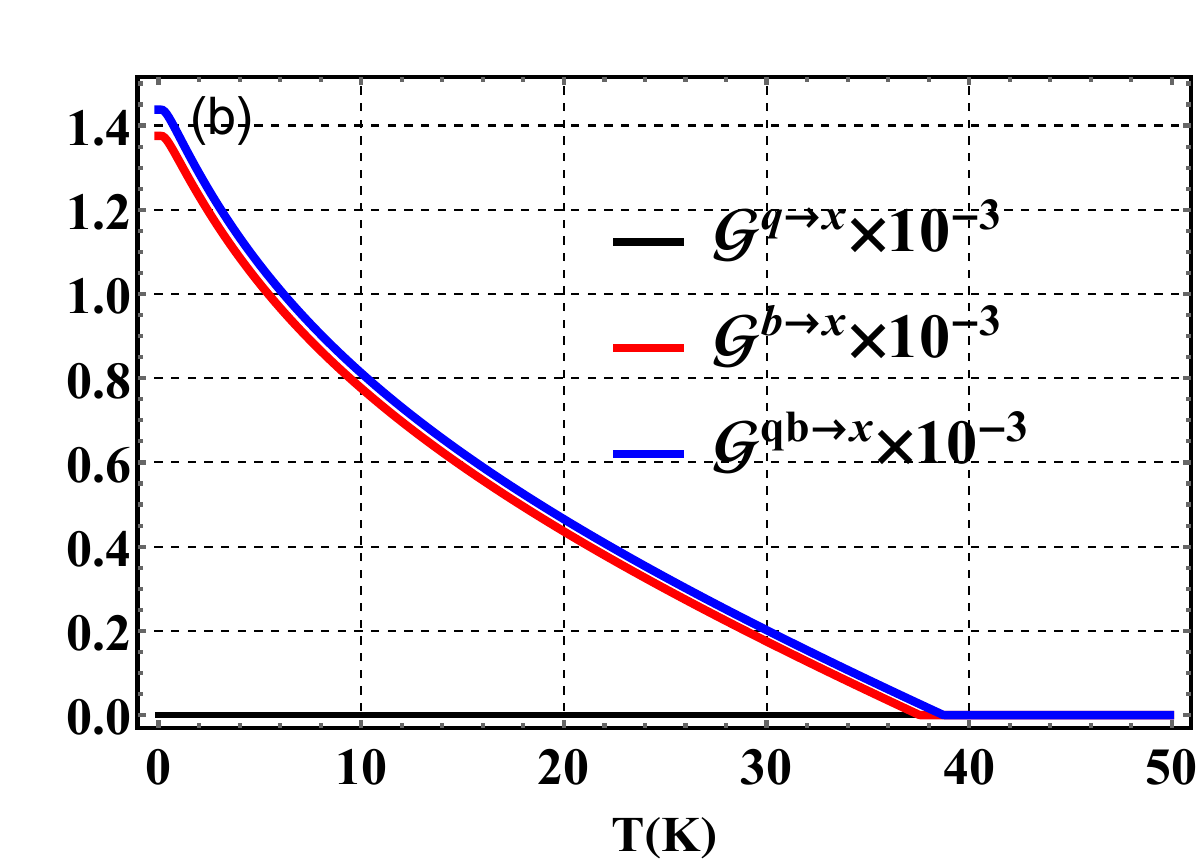}  
\end{tabular}
\caption{Plot the steady-state of the steerability as a function of temperature $\rm T$.  (a)
steering from the phonon to the exciton ($\mathcal{G}^{\rm b\rightarrow x}$), photon ($\mathcal{G}^{\rm b\rightarrow a}$), and cavity-exciton system ($\mathcal{G}^{\rm b\rightarrow ax}$); (b) steering from the photon to the exciton
($\mathcal{G}^{\rm a\rightarrow x}$), from the phonon to the exciton ($\mathcal{G}^{\rm b\rightarrow x}$), and from photon-phonon to exciton system ($\mathcal{G}^{\rm ab \rightarrow x}$), with detuning $\Delta_{\rm a}=-\Delta_{\rm x}={\bm \omega}_{\rm b}$.}
\label{fig2}
\end{figure} 
The results presented in Fig.~\ref{fig2} show that temperature has a strong detrimental effect on the Gaussian steerability of the system. As shown in Fig.~\ref{fig2}(a), both the steering from the phonon mode to the exciton mode, $\mathcal{G}^{\rm b\rightarrow x}$, and the collective steering to the joint exciton--photon subsystem, $\mathcal{G}^{\rm b\rightarrow xa}$, decrease as the temperature increases. This behavior is mainly due to the increasing thermal fluctuations, which introduce additional noise into the mechanical mode and progressively weaken the quantum correlations required for steering. At low temperatures, these correlations are better preserved, resulting in stronger steerability.

An important feature is that $\mathcal{G}^{\rm b\rightarrow xa}\geq\mathcal{G}^{\rm b\rightarrow x}$ over the considered temperature range. Thus, including the photon mode in the steered subsystem enhances the steering compared with the exciton mode alone. This collective effect makes the multipartite steering more robust against thermal noise. For the parameters considered here, the bipartite steering $\mathcal{G}^{\rm b\rightarrow x}$ eventually vanishes as the temperature increases, whereas the collective steering $\mathcal{G}^{\rm b\rightarrow xa}$ persists over a wider temperature range.

Figure~\ref{fig2}(b) leads to a similar conclusion. Both $\mathcal{G}^{\rm b\rightarrow x}$ and $\mathcal{G}^{\rm ab\rightarrow x}$ decrease with increasing temperature, reflecting the progressive degradation of the quantum correlations by thermal noise. At sufficiently low temperatures, the mechanical and exciton modes retain strong enough correlations to support Gaussian steering. As the temperature rises, the increasing thermal occupation of the mechanical mode introduces excess noise and eventually suppresses the steering. In contrast, no appreciable steering from the photon mode to the exciton mode is observed over the considered temperature range, indicating that the photon--exciton correlations alone are insufficient to generate Gaussian steering for the chosen parameters.

Finally, the results are consistent with the CKW-type monogamy relations for Gaussian steering. In particular, the collective steering satisfies
$
\mathcal{G}^{\rm b\rightarrow xa}>
\mathcal{G}^{\rm b\rightarrow x}+
\mathcal{G}^{\rm b\rightarrow a},
$
while
$
\mathcal{G}^{\rm ab\rightarrow x}>
\mathcal{G}^{\rm a\rightarrow x}+
\mathcal{G}^{\rm b\rightarrow x}.
$
These inequalities show that the collective steering exceeds the sum of the corresponding pairwise contributions, highlighting the monogamous distribution of Gaussian steering in the considered tripartite hybrid exciton--optomechanical system.
\\

\subsection{Dynamical state}
\begin{figure}[ht!]
\centering
\begin{tabular}{ccc}
\includegraphics[scale=0.4]{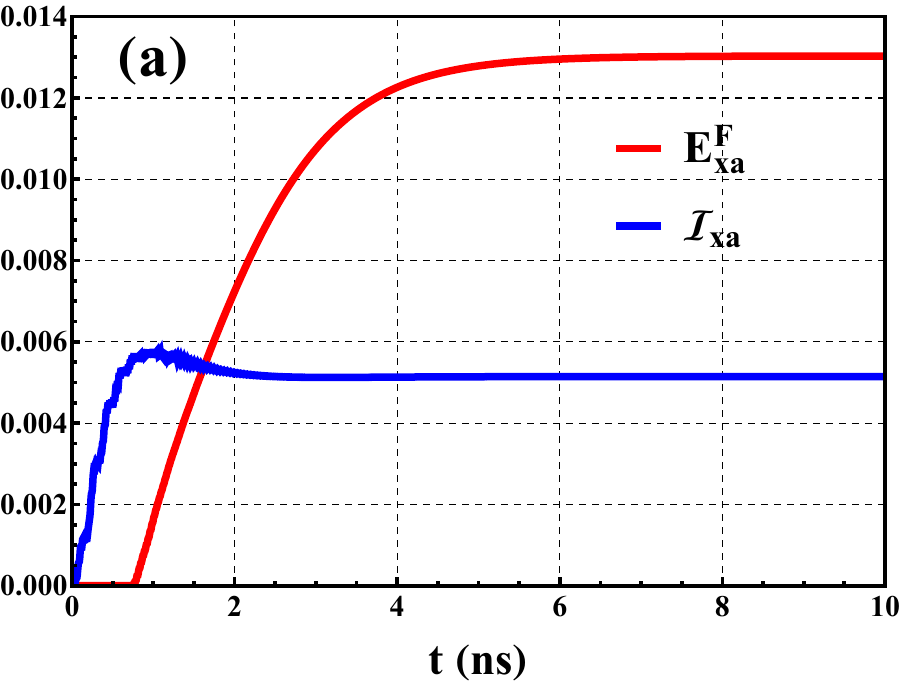} 
\includegraphics[scale=0.4]{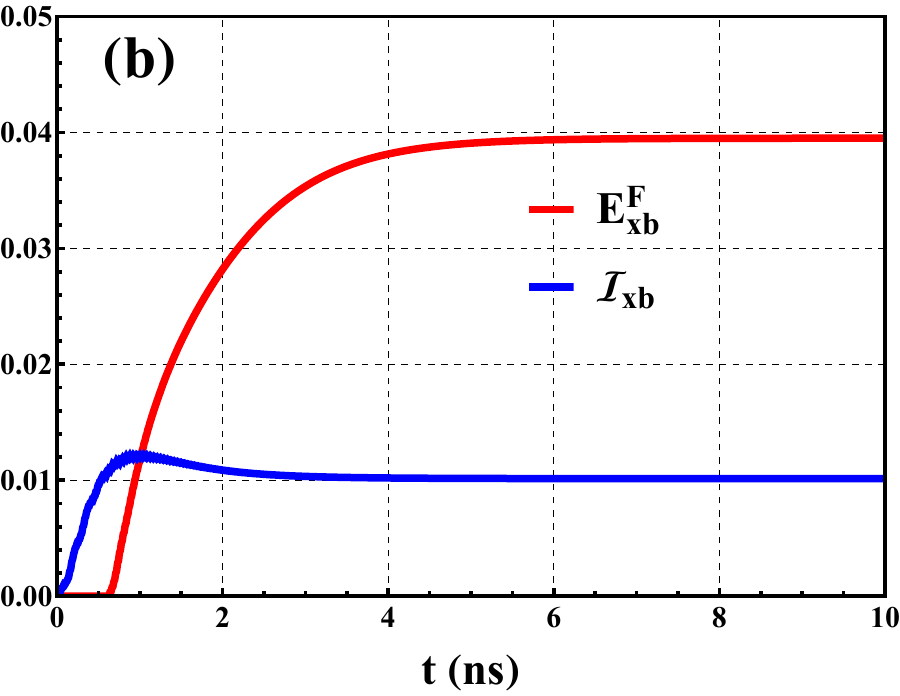} 
\includegraphics[scale=0.4]{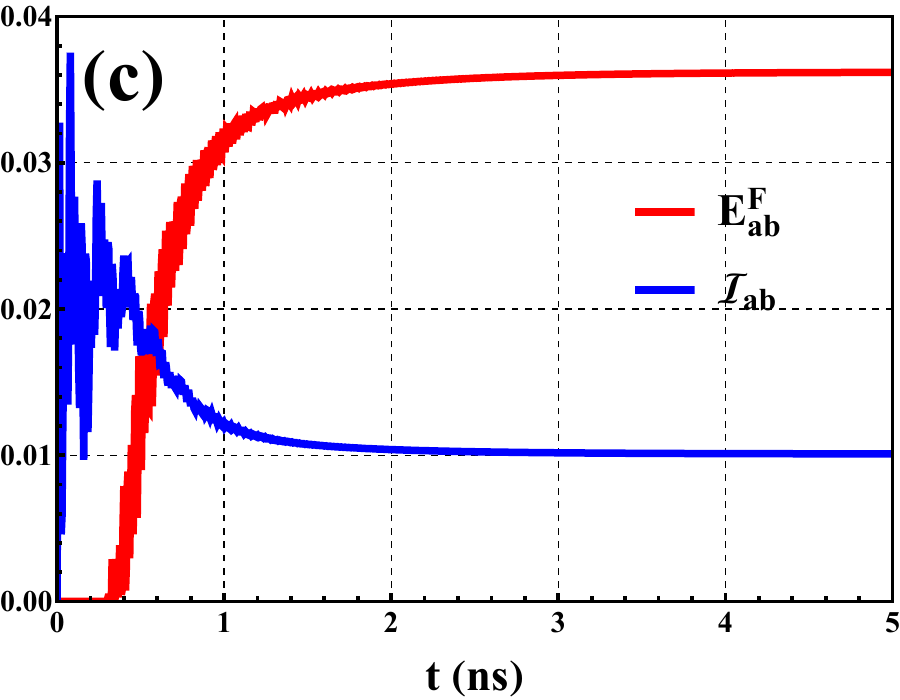} 
\end{tabular}
\caption{Plot the dynamic-state of the entanglement of formation and Gaussian interfometric power for (a) exciton--photon mode (${\rm E^{F}_{xa}},{\cal{I}_{\rm xa}}$), (b) exciton--phonon mode (${\rm E^{F}_{xb}},{\cal{I}_{\rm xb}}$) and (c) photon--phonon mode (${\rm E^{F}_{ab}},{\cal{I}_{\rm ab}}$), with $\Delta_{\rm a}=0.9{\bm \omega}_{\rm b}$and $\Delta_{\rm x}=-1.1{\bm \omega}_{\rm b}$.}
\label{fig8}
\end{figure}
In Fig.~\ref{fig8}, we investigate the time evolution of the entanglement of formation and Gaussian interferometric power for three bipartite subsystems: (a) exciton--photon mode (${\rm E^{F}_{xa}},{\cal{I}_{\rm xa}}$), (b) exciton--phonon mode (${\rm E^{F}_{xb}},{\cal{I}_{\rm xb}}$) and (c) photon--phonon mode (${\rm E^{F}_{ab}},{\cal{I}_{\rm ab}}$). This analysis allows us to compare the dynamical generation of entanglement and interferometric quantum resources, as well as their respective relaxation times toward the steady state.

As shown in Fig.~\ref{fig8}(a), the Eof of the exciton--photon subsystem (${\rm E}^{\rm F}_{\rm xa}$), is initially zero, indicating that the two modes are initially separable. For times ${\rm t}>0.5\,\mathrm{ns}$, the Eof starts to increase, revealing the gradual establishment of quantum entanglement between the excitonic and photonic modes. It subsequently approaches its steady-state value at approximately ${\rm t}=5\,\mathrm{ns}$, beyond which its variation becomes negligible. This behavior reflects the finite time required for the interaction dynamics to generate and stabilize bipartite entanglement. In contrast, the GIP becomes nonzero at an earlier stage, before the onset of finite entanglement. This observation indicates that the two modes can possess interferometric quantum resources even when their state remains separable. The GIP subsequently increases and eventually reaches a nearly constant value, signaling the stabilization of the interferometric properties of the subsystem.

A similar dynamical behavior is observed for the exciton--phonon subsystem in Fig.~\ref{fig8}(b). The Eof, ${\rm E}^{\rm F}_{\rm xb}$, initially vanishes and starts to increase for ${\rm t}>0.5\,\mathrm{ns}$, eventually reaching its steady-state value at approximately ${\rm t}=5\,\mathrm{ns}$. Interestingly, we find that ${\rm E}^{\rm F}_{\rm xb}>{\rm E}^{\rm F}_{\rm xa}$, indicating that the exciton--phonon interaction generates stronger bipartite entanglement than the exciton--photon interaction under the considered parameters. As in the previous case, the GIP becomes nonzero before the emergence of entanglement and subsequently increases until it reaches a nearly stationary value. This behavior further demonstrates that interferometric quantum resources can develop independently of the presence of bipartite entanglement. Overall, the exciton--photon and exciton--phonon subsystems exhibit similar relaxation dynamics, although the latter supports a higher steady-state entanglement.

For, Fig.~\ref{fig8}(c) illustrates the temporal evolution of the Eof and GIP for the photon--phonon subsystem. In this case, the Eof is initially zero and starts to increase for ${\rm t}>0.4\,\mathrm{ns}$, indicating the dynamical generation of entanglement between the photonic and mechanical modes. Interestingly, unlike the two previous subsystems, the ${\rm E^{F}_{ab}},{\cal{I}_{\rm ab}}$ reaches its steady-state value relatively rapidly, at approximately ${\rm t}=2\,\mathrm{ns}$. This faster convergence suggests that the photon--phonon subsystem establishes its stationary entangled state on a shorter timescale than the exciton--photon and exciton--phonon pairs. Meanwhile, the GIP is already nonzero during the initial separable stage and increases toward its maximum value, after which it becomes approximately time-independent. Thus, the interferometric quantum resource is established before the onset of entanglement and subsequently stabilizes as the system approaches its steady state.

Overall, these results reveal distinct dynamical timescales for the generation and stabilization of quantum resources in the three subsystems. While the exciton--photon and exciton--phonon pairs require approximately $5\,\mathrm{ns}$ to reach their steady-state entanglement, the photon--phonon pair reaches this regime within approximately $2\,\mathrm{ns}$. Furthermore, the earlier emergence of GIP compared with Eof in all three subsystems highlights the ability of Gaussian interferometric power to characterize nonclassical interferometric resources even in the absence of bipartite entanglement.\\

\begin{figure}[ht!]
\centering
\begin{tabular}{ccc}
\includegraphics[scale=0.4]{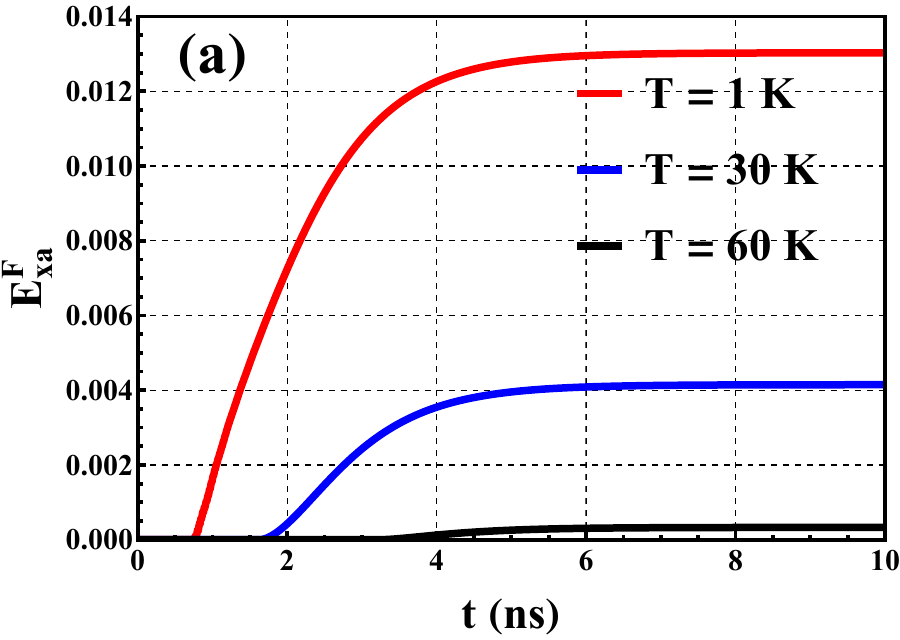} 
\includegraphics[scale=0.4]{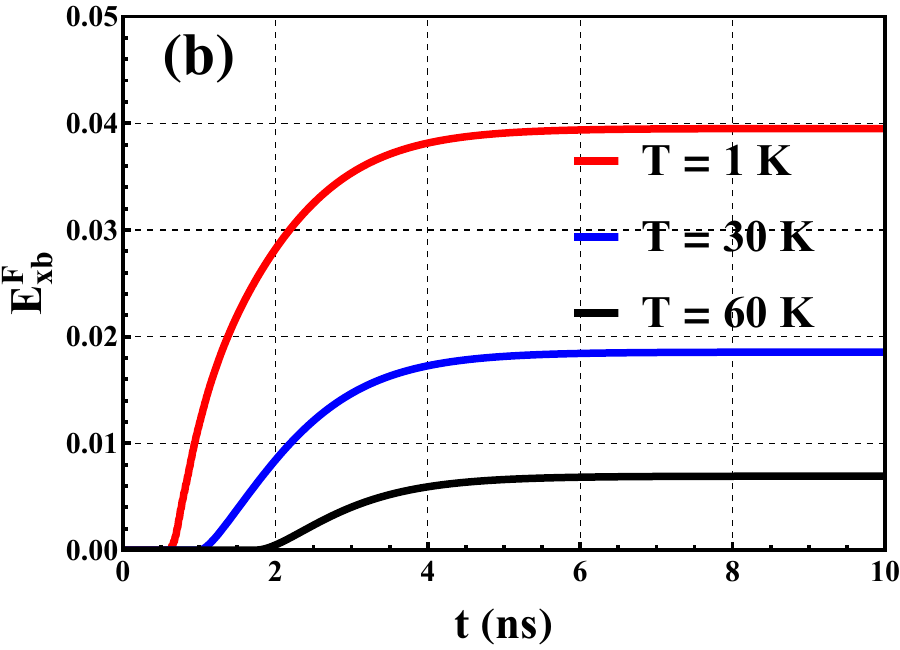} 
\includegraphics[scale=0.4]{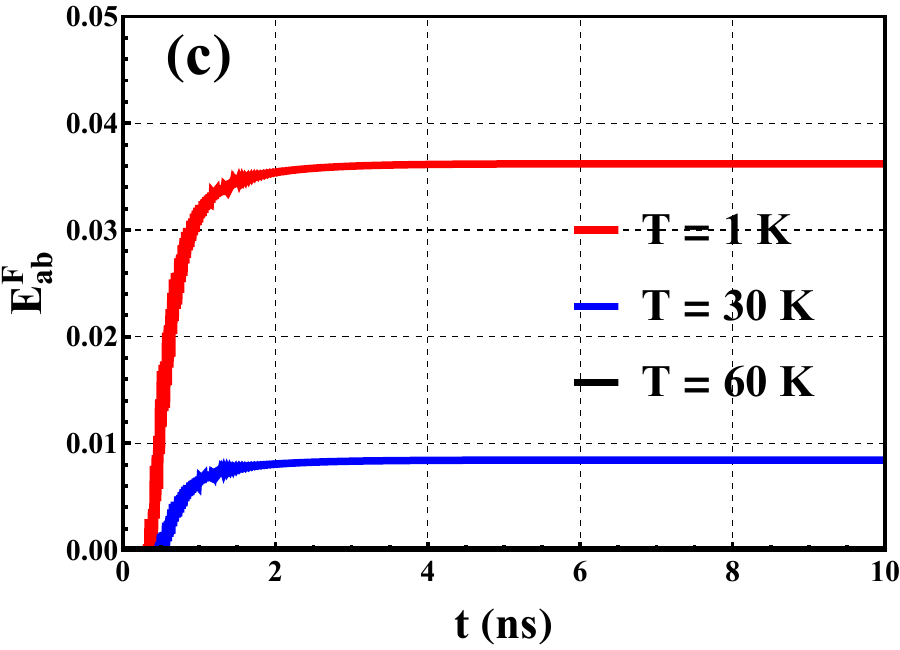} \\
\includegraphics[scale=0.4]{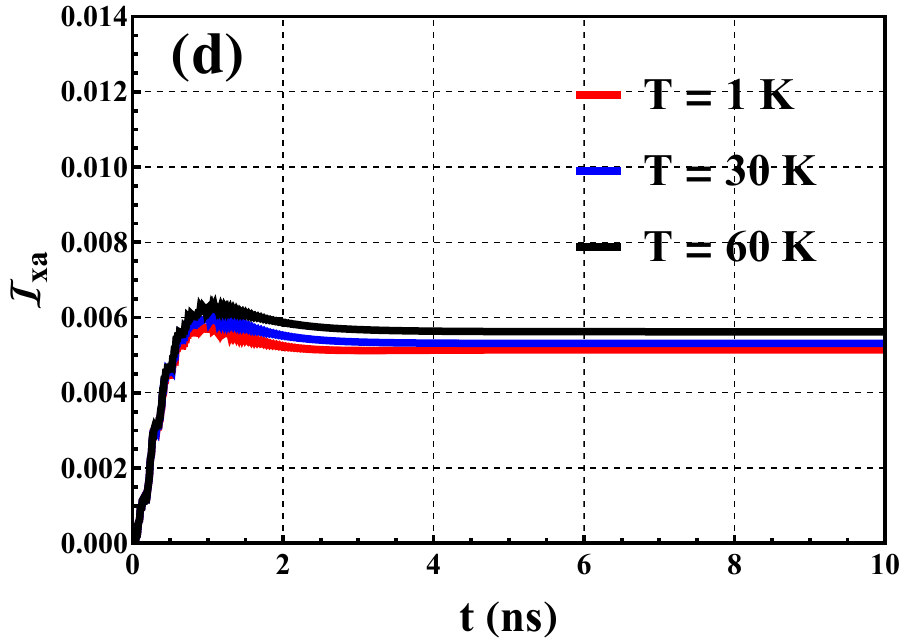}
\includegraphics[scale=0.4]{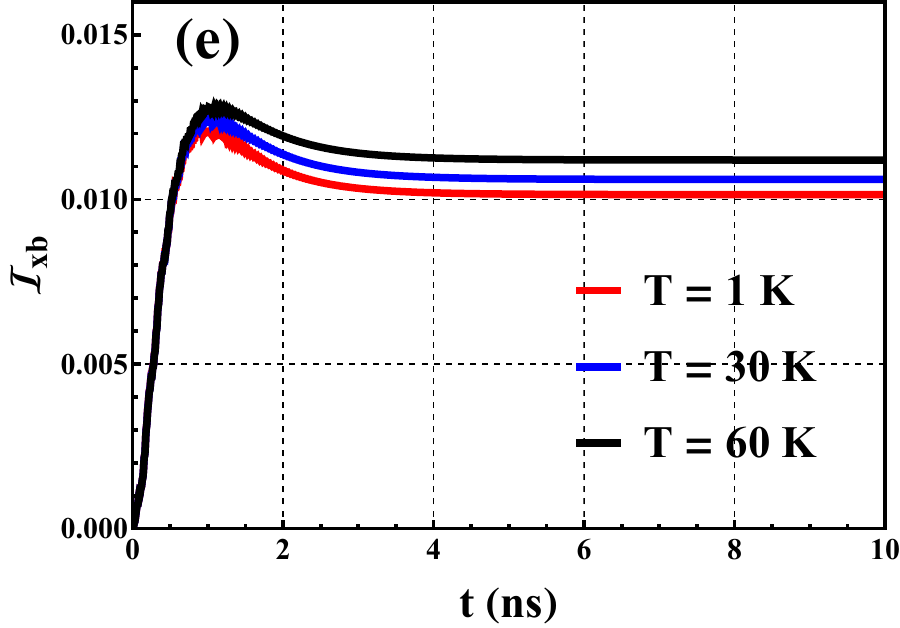}  
\includegraphics[scale=0.4]{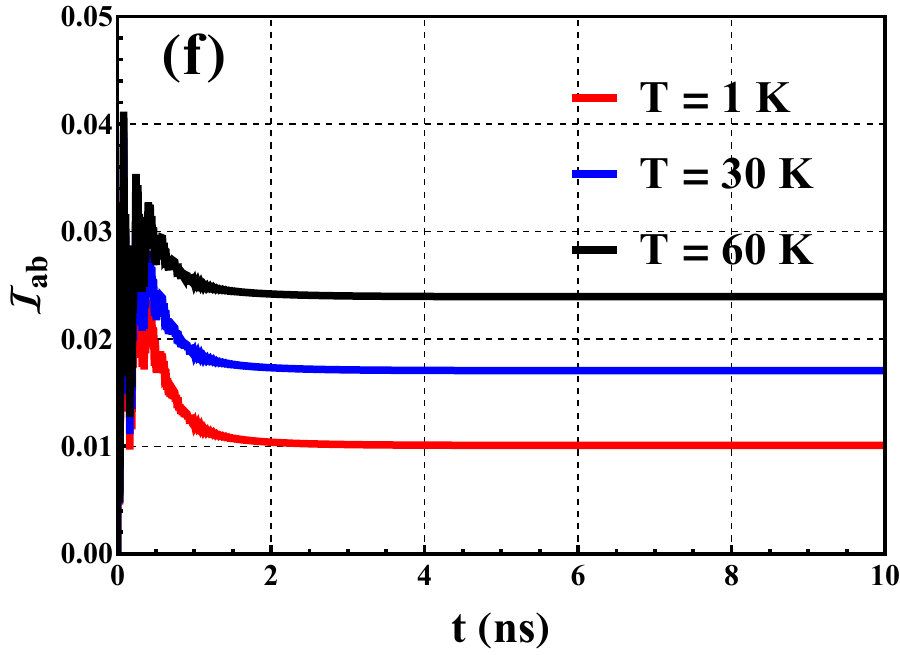} 
\end{tabular}
\caption{Plot the time evolution of (a–c) the entanglement of formation, and (d–f) the Gaussian interferometric power, for different bipartite subsystems: (a,d) exciton–photon (${\rm E^{F}_{xa}},{\cal{I}_{\rm xa}}$), (b,e) exciton–phonon (${\rm E^{F}_{xb}},{\cal{I}_{\rm xb}}$), and (c,f) photon–phonon modes (${\rm E^{F}_{ab}},{\cal{I}_{\rm ab}}$), for various values of the temperature \(\rm T\). The other parameters are fixed at \(\Delta_{\rm a}=0.9\omega_{\rm b}\) and \(\Delta_{\rm x}=-1.1\omega_{\rm b}\).}
\label{fig9}
\end{figure}   

In Fig.~\ref{fig9}, we investigate the time evolution of the entanglement of formation and Gaussian interferometric power for three different bipartite subsystems: (a,d) exciton--photon (${\rm E}^{F}_{xa},{\cal I}_{\rm xa}$), (b,e) exciton--phonon (${\rm E}^{F}_{xb},{\cal I}_{\rm xb}$), and (c,f) photon--phonon (${\rm E}^{F}_{ab},{\cal I}_{\rm ab}$), for different values of the temperature ${\rm T}$. This analysis allows us to explore how thermal fluctuations influence the transient dynamics and steady-state behavior of entanglement and interferometric quantum correlations in the hybrid system.

As shown in Fig.~\ref{fig9}(a)--(c), the Eof initially remains zero over a finite time interval, indicating that the corresponding bipartite subsystems are initially in separable states. The duration of this initial entanglement-free interval increases with temperature, revealing that thermal fluctuations delay the generation of bipartite entanglement. Furthermore, increasing the temperature progressively reduces the Eof, indicating that thermal noise weakens the quantum entanglement established between the different pairs of modes. This behavior is physically expected because thermal excitations introduce additional fluctuations into the system, thereby degrading the quantum correlations generated through the coherent interactions. A comparison of the three subsystems also reveals that the exciton--photon entanglement is considerably weaker than that of the other two pairs, suggesting that the effective interaction pathways favor the generation of entanglement in the exciton--phonon and photon--phonon subsystems. In particular, the photon--phonon pair exhibits a comparatively stronger entanglement, highlighting the important role of the optomechanical interaction in establishing quantum correlations between these two modes.

In contrast, Fig.~\ref{fig9}(d)--(f) reveals a different dynamical and thermal behavior of the GIP. For all three bipartite subsystems, the GIP gradually increases from its initial value and approaches a steady-state regime for $t>2~\rm ns$. This temporal evolution indicates that the establishment of interferometric quantum correlations requires a finite interaction time before the system reaches its asymptotic regime. Interestingly, unlike the Eof, the GIP exhibits a slight enhancement with increasing temperature. This counter intuitive behavior suggests that thermal fluctuations do not necessarily degrade all forms of quantum correlations in the same manner. Although thermal noise suppresses entanglement, it can modify the local and intermodal fluctuations in such a way that the sensitivity of the subsystem to phase encoding, as quantified by the GIP, is slightly enhanced. Consequently, the GIP can remain finite or even increase in parameter regimes where the Eof is strongly suppressed. This distinction emphasizes that interferometric power captures quantum correlations relevant to phase-estimation tasks rather than entanglement alone. Moreover, the GIP associated with the exciton--phonon and photon--phonon pairs is larger than that of the exciton--photon pair, indicating that these two subsystems exhibit a stronger sensitivity to local phase encoding.

Overall, these results demonstrate that entanglement and Gaussian interferometric power exhibit markedly different dynamical responses to thermal fluctuations. While increasing temperature delays entanglement generation and reduces its magnitude, the GIP displays a slight thermal enhancement and approaches its steady-state value on a relatively short timescale. These contrasting trends highlight the complementary nature of the two quantifiers: entanglement characterizes the nonseparability of quantum states, whereas the GIP characterizes their usefulness for quantum phase estimation. Importantly, the persistence of GIP in regimes where entanglement is weak suggests that useful interferometric quantum correlations may survive even when bipartite entanglement is strongly degraded by thermal noise.\\

\begin{figure}[ht!]
\centering
\begin{tabular}{ccc}
\includegraphics[scale=0.45]{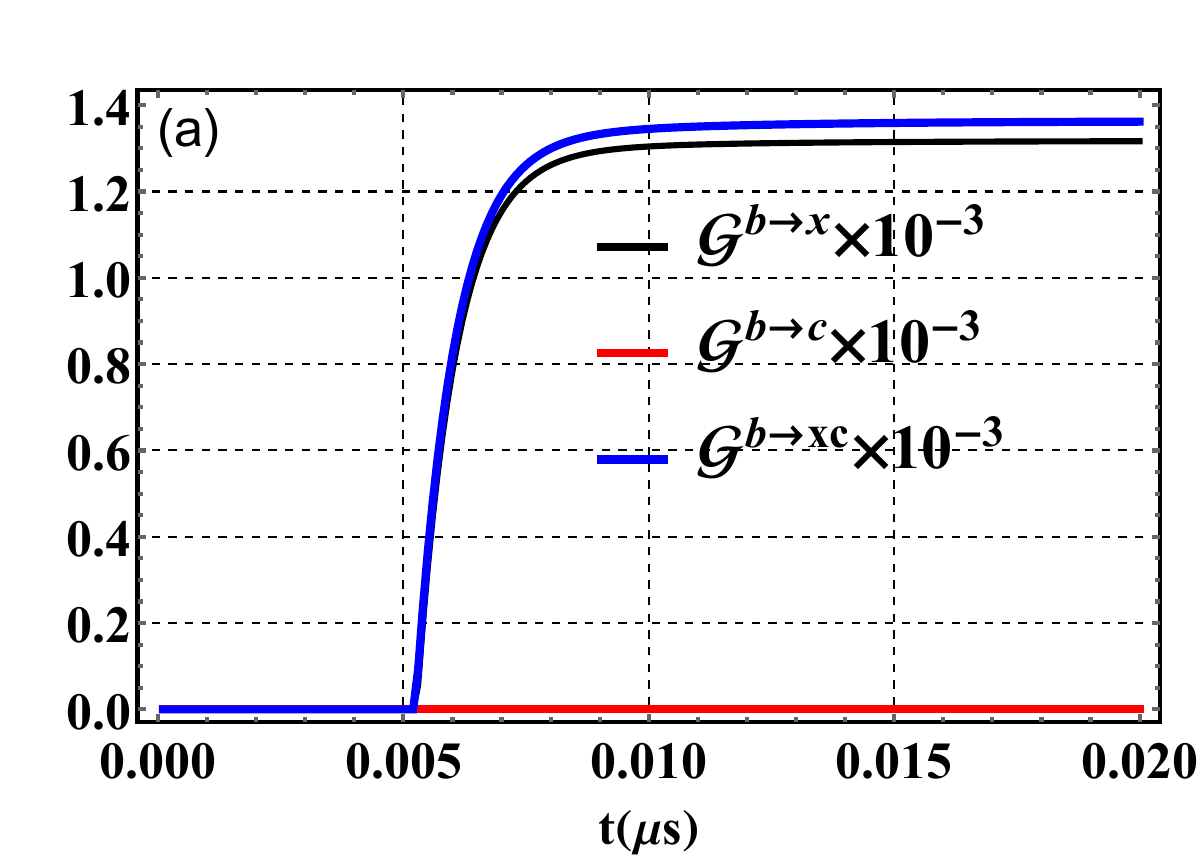} 
\includegraphics[scale=0.45]{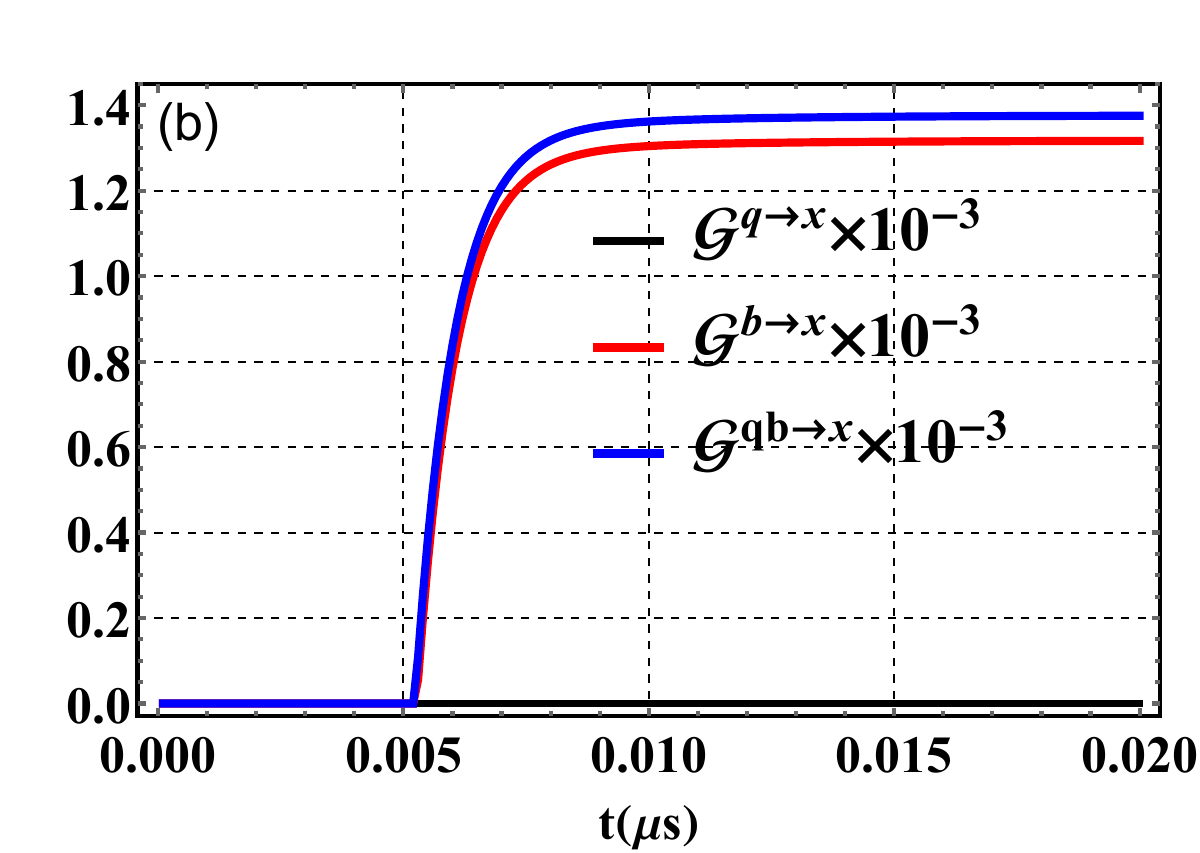} 
\end{tabular}
\caption{Plot the dynamic-state of the Gaussian quantum steering, (a)
steering from the phonon to the exciton ($\mathcal{G}^{\rm b\rightarrow x}$), photon ($\mathcal{G}^{\rm b\rightarrow a}$), and cavity-exciton system ($\mathcal{G}^{\rm b\rightarrow ax}$); (b) steering from the photon to the exciton
($\mathcal{G}^{\rm a\rightarrow x}$), from the phonon to the exciton ($\mathcal{G}^{\rm b\rightarrow x}$), and from photon-phonon to exciton system ($\mathcal{G}^{\rm ba \rightarrow x}$), with detuning $\Delta_{\rm a}=-\Delta_{\rm x}={\bm \omega}_{\rm b}$.}
\label{fig7}
\end{figure} 
In Fig. \ref{fig7}, we investigate the dynamical evolution of the Gaussian steering between the different modes of the proposed hybrid system. This analysis provides insight into the time required for quantum correlations to develop and allows us to characterize the robustness and strength of the resulting steering during the temporal evolution.

As shown in Fig. \ref{fig7}(a), the Gaussian steering from the phonon mode toward the exciton mode, $\rm b\rightarrow x$, and toward the joint exciton--photon subsystem, $\rm b\rightarrow xa$, does not emerge instantaneously. Instead, the steering becomes appreciable only after a finite interaction time, approximately ${\rm t}>0.005~\mu$s. This delayed onset indicates that the quantum correlations responsible for Gaussian steering require a finite time to build up through the interactions between the different modes. In particular, we find that the collective steering satisfies $\mathcal{G}^{\rm b\rightarrow xa}>\mathcal{G}^{\rm b\rightarrow x}$, demonstrating that the phonon mode can steer the joint exciton--photon subsystem more strongly than the exciton mode alone. This enhancement highlights the role of the photon mode in supporting and redistributing the quantum correlations within the hybrid system. On the other hand, the quantity $\mathcal{G}^{\rm b\rightarrow x}$ remains zero throughout the entire evolution, indicating that no Gaussian steering is established from the phonon mode to the exciton mode for the considered parameters.

Furthermore, Fig.\ref{fig7}(b) presents the time evolution of the steering quantities $\mathcal{G}^{\rm b\rightarrow x}$ and $\mathcal{G}^{\rm ab\rightarrow x}$. Similar to Fig. \ref{fig4}(a), the steering correlations develop only after a finite transient period, with a noticeable onset around ${\rm t}>0.005~\mu$s. This behavior confirms that the generation of Gaussian steering is a dynamical process rather than an instantaneous effect. We also observe that $\mathcal{G}^{\rm ab\rightarrow x}>\mathcal{G}^{\rm b\rightarrow x}$, meaning that the joint photon--phonon subsystem can steer the exciton mode more efficiently than the phonon mode alone. The additional photon degree of freedom therefore enhances the transfer and preservation of quantum correlations toward the exciton mode. In contrast, Gaussian steering between the photon and exciton modes alone is absent throughout the entire time evolution, indicating that their direct correlations remain below the threshold required to generate Gaussian steerability.

Overall, the results shown in Fig.~\ref{fig4} highlight the important role of temporal evolution in the generation and redistribution of quantum steering. The finite onset time can be attributed to the time required for the intermodal quantum correlations to build up. Moreover, the stronger steering observed for the collective quantities $\mathcal{G}^{\rm b\rightarrow xa}$ and $\mathcal{G}^{\rm ab\rightarrow x}$ shows that including an additional mode can significantly enhance the steerability of the system. This behavior indicates that multimode correlations can be beneficial for generating and maintaining robust Gaussian steering in hybrid exciton--optomechanical systems.

The results also show that the considered Gaussian state satisfies the CKW-type monogamy relations for steering. In particular, the collective steering from the mechanical mode to the joint exciton--photon subsystem obeys
$
\mathcal{G}^{\rm b\rightarrow xa} >
\mathcal{G}^{\rm b\rightarrow x}
+\mathcal{G}^{\rm b\rightarrow a},
$
as shown in Fig.~\ref{fig4}(a). Likewise, the steering from the joint exciton--mechanical subsystem to the exciton mode satisfies
$
\mathcal{G}^{\rm ab\rightarrow x} >
\mathcal{G}^{\rm a\rightarrow x}
+\mathcal{G}^{\rm b\rightarrow x},
$
as shown in Fig.~\ref{fig4}(b). These inequalities demonstrate that the collective steering can exceed the sum of the corresponding pairwise contributions, confirming the monogamous distribution of Gaussian steering in the considered tripartite system.
\section{Experimental feasibility}
\label{sec5}
The hybrid exciton--optomechanical system that is discussed in this study can be realized using proven techniques for cavity optomechanics in semiconductors.  The system consists of a semiconductor quantum well (QW) embedded in an optical microcavity formed by two distributed Bragg reflectors (DBRs). In the strong exciton--photon coupling regime, the excitonic mode interacts coherently with the cavity photon mode, a regime that has been experimentally demonstrated in semiconductor microcavities and has led to the observation of exciton-polariton states. Such platforms provide exciton--photon coupling strengths in the GHz range while maintaining sufficiently low dissipation rates to preserve coherent light--matter interactions.

The mechanical degree of freedom is provided by the movable DBRs, whose motion couples to the intracavity optical field through radiation pressure. This type of optomechanical interaction has been experimentally demonstrated in a variety of cavity-optomechanical platforms. In the present configuration, the mechanical motion is therefore coupled to the cavity field, while the cavity field is simultaneously coupled to the exciton mode. Under strong coherent driving, the optomechanical interaction can be enhanced through linearization, providing an effective mechanism for generating and transferring quantum correlations among the exciton, optical, and mechanical modes.

The experimental characterization of the resulting Gaussian state requires the reconstruction of its first-order moments and covariance matrix. The optical fluctuations can be directly accessed by measuring the output field of the cavity using standard homodyne or heterodyne detection techniques. Homodyne detection provides access to a selected field quadrature through interference with a phase-controlled local oscillator, whereas heterodyne detection enables the simultaneous acquisition of both field quadratures by using a frequency-shifted local oscillator. These well-established techniques provide the required statistical information for the characterization of continuous-variable Gaussian states \cite{Yadin2018,Yuen1983}.

In contrast, the exciton mode cannot be directly detected using conventional optical homodyne or heterodyne measurements. Therefore, in the proposed detection scheme, the exciton information is transferred to an auxiliary optical probe cavity. The exciton mode is coupled to this auxiliary cavity with coupling strength ${\rm g}_{\rm xc}$, while the auxiliary cavity is driven by a weak blue-detuned probe field. This configuration enables the excitonic fluctuations to be mapped onto the optical output field of the auxiliary cavity, thereby providing an experimentally accessible readout of the exciton state.

\begin{figure}[ht!]
\centering
\includegraphics[scale=0.5]{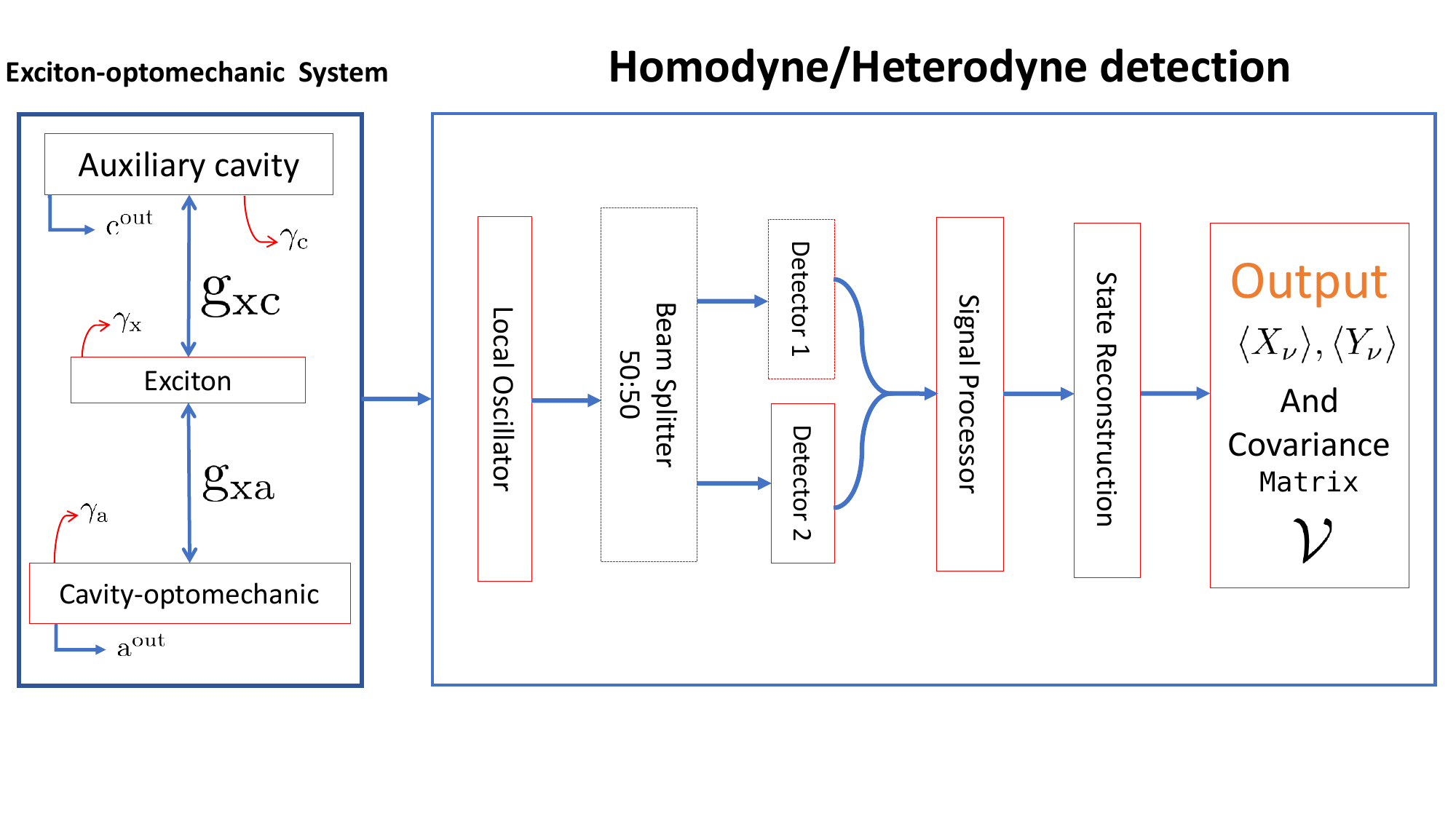}
\caption{Schematic illustration of the reconstruction of the first- and second-order moments of the Gaussian state using homodyne and heterodyne detection. Here, ${\rm g}_{\rm xc}$ denotes the coupling strength between the auxiliary cavity and the exciton mode. The reconstruction relies on the input--output relations $
{\rm a}^{\rm out}={\rm a}^{\rm in}-\sqrt{2\gamma_{\rm a}}\,{\rm a},$
${\rm c}^{\rm out}={\rm c}^{\rm in}-\sqrt{2\gamma_{\rm c}}\,{\rm c},$
where ${\rm c}^{\rm in}$ represents the input field of the auxiliary cavity and $\gamma_{\rm c}$ is its corresponding decay rate.}
\label{fig12}
\end{figure}

The corresponding output field is continuously monitored, and the quadrature fluctuations are recorded as time-dependent measurement traces. From these traces, the probability distributions of the measured quadratures can be reconstructed, allowing the first-order moments of the Gaussian state to be determined. The covariance matrix is subsequently obtained from the second-order correlations between the measured quadratures. In particular, the experimentally recorded traces provide the pairwise correlations and ensemble averages required to reconstruct all elements of the covariance matrix. Consequently, the combination of the auxiliary-cavity readout with homodyne or heterodyne detection provides access to the complete first- and second-order moments of the exciton--photon--mechanical Gaussian state, which are the quantities required for evaluating the quantum correlations and multiparameter estimation bounds investigated in this work. Figure~\ref{fig12} illustrates the proposed homodyne and heterodyne detection scheme employed to characterize the Gaussian state by measuring its first-order moments and covariance matrix.
\section{Conclusion}   \label{sec6}

In conclusion, we have investigated quantum correlations in a hybrid exciton--cavity optomechanical system, considering both steady-state and dynamical regimes. In particular, we have analyzed the effects of exciton-drive detuning, input power, and thermal noise on the generation and evolution of quantum correlations, quantified through the Eof, GIP, and Gaussian quantum steering. Our results reveal that these quantum measures exhibit distinct behaviors as the system parameters vary. In particular, the GIP remains finite over a broader parameter range than the Eof, demonstrating that quantum correlations detectable through interferometric power can persist even in the absence of bipartite entanglement. Furthermore, we find that the Eof of the exciton--phonon subsystem exhibits greater robustness than that of the other considered bipartite subsystems, highlighting the important role of exciton--phonon interactions in establishing and preserving entanglement.
Regarding Gaussian steering, our results demonstrate that thermal fluctuations generally have a detrimental effect on steerability. Specifically, increasing the temperature progressively suppresses Gaussian steering by introducing additional thermal noise and weakening the quantum correlations shared among the different modes. Moreover, thermal fluctuations can modify the directional properties of steering, leading to regimes in which one-way steering persists even when two-way steering is no longer possible.
In the dynamical regime, we show that Gaussian steering can emerge and become enhanced during the temporal evolution of the system, highlighting the importance of transient dynamics in generating and distributing quantum correlations. The time-dependent behavior of the quantum measures therefore provides additional opportunities for controlling the steering properties beyond those available in the steady-state regime. 
Finally, we have investigated the distribution of Gaussian steering among the three modes by analyzing the corresponding CKW-type monogamy inequalities. Our results confirm that these monogamy relations are satisfied throughout the investigated parameter regimes, demonstrating the constraints they impose on the distribution of bipartite steering within the tripartite system. Overall, our findings establish the hybrid exciton--optomechanical platform as a promising candidate for the generation, characterization, and control of multipartite quantum resources, with potential applications in quantum information processing.

\end{document}